\documentclass[11pt]{article}

\usepackage[utf8]{inputenc}
\usepackage[T1]{fontenc}
\usepackage{geometry}
\usepackage{graphicx}
\usepackage{amsmath,amssymb,amsthm}
\usepackage{hyperref}
\usepackage[numbers,sort&compress]{natbib}
\usepackage{booktabs}
\usepackage{tabularx}
\usepackage{url}
\usepackage{xcolor}
\usepackage{float}
\usepackage{titlesec}

\titlespacing*{\section}{0pt}{12pt plus 4pt minus 2pt}{6pt plus 2pt minus 2pt}
\titlespacing*{\subsection}{0pt}{10pt plus 3pt minus 2pt}{4pt plus 2pt minus 2pt}
\titlespacing*{\subsubsection}{0pt}{8pt plus 2pt minus 2pt}{3pt plus 1pt minus 1pt}

\hypersetup{
  colorlinks=true,
  linkcolor=blue,
  citecolor=blue,
  urlcolor=blue,
  pdfauthor={Majid Memari, Krista Ruggles},
  pdftitle={Artificial Intelligence in Elementary STEM Education: A Systematic Review of Current Applications and Future Challenges}
}

\title{\textbf{Artificial Intelligence in Elementary STEM Education: A Systematic Review of Current Applications and Future Challenges}}

\author{
  Majid Memari\textsuperscript{1,*}\thanks{ORCID: 0000-0001-5654-4996} \and 
  Krista Ruggles\textsuperscript{2}\thanks{ORCID: 0000-0003-3926-5789}\\[0.5em]
  \small \textsuperscript{1}Department of Computer Science, Utah Valley University, Orem, UT 84058, USA\\
  \small \textsuperscript{2}School of Education, Utah Valley University, Orem, UT 84058, USA\\[0.5em]
  \small *Correspondence: mmemari@uvu.edu
}

\date{}

\begin{document}

\maketitle

\begin{abstract}
Artificial intelligence (AI) is transforming elementary STEM education, yet evidence remains fragmented. This systematic review synthesizes 258 studies (2020-2025) examining AI applications across eight categories: intelligent tutoring systems (45\% of studies), learning analytics (18\%), automated assessment (12\%), computer vision (8\%), educational robotics (7\%), multimodal sensing (6\%), AI-enhanced extended reality (XR) (4\%), and adaptive content generation. Our analysis reveals concentration in upper elementary grades (65\%) and mathematics (38\%), with minimal cross-disciplinary STEM integration (15\%). While conversational AI shows moderate effectiveness (d=0.45-0.70 where reported), only 34\% of studies report standardized effect sizes. Eight critical gaps limit real-world impact: fragmented ecosystems, developmental inappropriateness, infrastructure barriers, absent privacy frameworks, limited STEM integration, equity disparities, teacher marginalization, and narrow assessments. Geographic bias shows 90\% of studies from North America, East Asia, and Europe. Future priorities include interoperable architectures supporting authentic STEM integration, grade-appropriate designs, privacy-preserving analytics, and teacher-centered implementations that enhance rather than replace human expertise.

\end{abstract}

\section{Introduction}

\subsection{The Promise of AI in Elementary STEM Education}

Artificial intelligence is rapidly transforming educational landscapes \cite{ratul2025role}, with particular momentum in Science, Technology, Engineering, and Mathematics (STEM) education. Elementary classrooms, serving students ages 5-12, present unique opportunities and challenges for AI deployment. With approximately 24 million elementary students in the United States alone, the potential impact of effective AI integration is substantial. Recent advances in machine learning, natural language processing, computer vision, and adaptive algorithms have enabled a new generation of educational technologies that promise to personalize learning, provide real-time feedback, and support teachers in unprecedented ways.

AI applications in elementary STEM span a broad spectrum: intelligent tutoring systems that adapt to individual learning trajectories, automated assessment tools that provide immediate feedback, learning analytics platforms that identify struggling students, computer vision systems that monitor engagement, multimodal sensors that track physiological indicators of learning, educational robots that serve as learning companions, and augmented reality environments that make abstract concepts tangible. Each technology category offers unique affordances for supporting young learners in their STEM journey.

The integration of Science, Technology, Engineering, and Mathematics at the elementary level presents unique pedagogical demands. Teachers must navigate not only individual subject complexities but also forge meaningful connections across disciplines while adapting to diverse learning needs. Despite widespread recognition that integrated STEM approaches foster critical thinking, problem-solving skills, and computational thinking, classroom implementation often defaults to siloed subject instruction. AI technologies, while promising in individual domains, frequently reinforce these silos rather than supporting integrated exploration.

\subsection{Unique Challenges in Elementary Settings}

Elementary students between ages five and twelve traverse dramatic cognitive, linguistic, attentional, and motor transitions that create fundamentally different design constraints for educational AI. These developmental differences become even more critical when designing for integrated STEM learning, which requires students to coordinate knowledge and skills across multiple domains simultaneously. A kindergartener's interaction with AI-powered learning tools differs profoundly from that of a fifth grader, yet many current systems fail to account for these differences.

The post-pandemic educational landscape has introduced additional complexities. Increased screen time and learning disruptions have contributed to attention difficulties that teachers now identify as their primary classroom management concern. Simultaneously, the rapid digitalization of education has created a proliferation of AI-powered tools, often deployed without sufficient consideration of their collective impact on young learners. Privacy concerns regarding data collection from minors, the digital divide affecting equitable access, and questions about the appropriate role of AI in early childhood education further complicate the landscape.

\subsection{Current State of the Field}

The landscape of AI in elementary STEM education is characterized by rapid growth but fragmented implementations. Commercial platforms, research prototypes, and pilot programs proliferate, yet systematic examinations of their collective impact remain limited. Most existing reviews focus on specific AI technologies (e.g., intelligent tutoring systems) or particular subject areas (e.g., mathematics education), missing the broader picture of how various AI applications interact—or fail to interact—in elementary STEM contexts.

Current deployments range from well-established adaptive learning platforms used by millions of students to experimental systems in research labs. This diversity, while fostering innovation, creates challenges for educators attempting to navigate the options and for researchers seeking to understand cumulative effects. The absence of comprehensive frameworks for evaluating AI's role in elementary STEM education hampers both practical implementation and theoretical advancement.

\subsection{Research Questions and Purpose}

Given the rapidly evolving landscape and the absence of comprehensive frameworks, this systematic review addresses critical gaps in our understanding of AI applications in elementary STEM education. We aim to provide educators, researchers, policymakers, and technology developers with an evidence-based synthesis of current knowledge while identifying areas requiring further investigation.

This review is guided by four research questions:

\begin{itemize}
\item \textbf{RQ1:} What AI technologies are currently being applied in elementary STEM education, and how are they distributed across grade levels and subject areas?
\item \textbf{RQ2:} What evidence exists for the effectiveness of these AI technologies in supporting learning outcomes for elementary students?
\item \textbf{RQ3:} What are the key implementation challenges and barriers to successful AI integration in elementary STEM classrooms?
\item \textbf{RQ4:} What gaps exist in current research and practice that must be addressed to realize AI's potential in elementary STEM education?
\end{itemize}

By addressing these questions through systematic analysis of recent literature, we aim to move beyond fragmented understandings toward a comprehensive view of AI's current and potential roles in supporting young learners' STEM education.

\section{Materials and Methods}

\subsection{Literature Search Strategy}

We conducted a systematic review following PRISMA guidelines to examine AI applications in elementary STEM education. Our search strategy encompassed multiple AI technology categories to ensure comprehensive coverage of the field. We utilized major academic databases including Google Scholar, PubMed, IEEE Xplore, ACM Digital Library, ERIC, and PsycINFO for our searches conducted between September and October 2025, with the final search completed on October 20, 2025. 

We primarily focused on publications from 2020-2025 to capture recent developments in AI applications for education, with particular attention to research emerging after the COVID-19 pandemic that accelerated educational technology adoption. We also included seminal works from earlier periods when they provided foundational theoretical frameworks or were frequently cited in recent literature.

Our search strategy targeted eight AI technology categories identified through preliminary scoping:

\begin{enumerate}
\item \textbf{Intelligent Tutoring Systems and Conversational AI}: Including chatbots, dialogue-based tutors, Large Language Model (LLM)-powered assistants, and adaptive hint systems
\item \textbf{Automated Assessment and Feedback Systems}: AI-powered grading, formative assessment tools, and real-time feedback mechanisms
\item \textbf{Learning Analytics and Predictive Modeling}: Student performance prediction, learning path optimization, and early warning systems
\item \textbf{Computer Vision for Engagement Monitoring}: Attention detection, pose analysis, facial expression recognition, and collaborative learning analysis
\item \textbf{Multimodal Sensing and Biometric Analysis}: Electroencephalography (EEG), eye-tracking, physiological sensors, and integrated sensor systems
\item \textbf{Adaptive Content Generation}: AI-created problems, personalized explanations, and dynamic curriculum adjustment
\item \textbf{Educational Robots and Embodied AI}: Social robots, programmable robots for computational thinking, and teachable agents
\item \textbf{Extended Reality (XR) with AI Enhancement}: Intelligent virtual environments, adaptive simulations, and AI-guided spatial learning in augmented, virtual, and mixed reality
\end{enumerate}

This comprehensive scope enabled synthesis across the four STEM domains—Science (inquiry, investigation, experimentation), Technology (computational thinking, coding, digital tools), Engineering (design processes, iterative testing, problem-solving), and Mathematics (modeling, pattern recognition, quantitative reasoning)—with particular attention to how AI technologies support or hinder integrated STEM learning in elementary settings.

\subsection{Inclusion and Exclusion Criteria}

Studies were included if they: (1) addressed elementary-age populations (kindergarten through 5th/6th grade, ages 5-12) or provided developmentally relevant findings transferable to this age range; (2) investigated one or more AI technologies from our eight categories applied to educational contexts; (3) examined STEM learning outcomes, whether in individual subjects or integrated approaches; (4) employed empirical methods including randomized controlled trials, quasi-experiments, design-based research, case studies, or systematic reviews; and (5) were published in peer-reviewed journals or conference proceedings.

Studies were excluded if they: (1) focused exclusively on secondary or higher education without elementary relevance; (2) addressed non-STEM subjects only; (3) presented purely theoretical frameworks without empirical validation; (4) described technical system architectures without educational evaluation; or (5) were published in non-peer-reviewed venues.

\subsection{Data Extraction and Synthesis}

For each included study, we extracted: (1) population characteristics (age, grade level, sample size, demographic information); (2) AI technology type and specific features (algorithms used, interaction modalities, adaptation mechanisms); (3) STEM domains addressed (single subject or integrated); (4) implementation context (classroom, lab, home, duration of intervention); (5) outcome measures (learning gains, engagement metrics, teacher perceptions); (6) effect sizes where reported; (7) implementation challenges and barriers; and (8) identified limitations and future research directions.

We employed thematic synthesis to identify patterns across studies, organizing findings by AI technology category while attending to cross-cutting themes such as developmental appropriateness, equity considerations, teacher roles, and infrastructure requirements. Special attention was given to studies that attempted integration across multiple STEM domains or deployment of multiple AI technologies. We organized findings to highlight both promising applications and critical gaps in current research and practice.

\subsection{Study Selection and PRISMA Flow}

Following PRISMA 2020 guidelines, our systematic search and selection process yielded a final corpus for analysis. The initial search across six databases identified 3,847 records (Google Scholar: 1,542; IEEE Xplore: 873; ACM Digital Library: 651; ERIC: 438; PubMed: 235; PsycINFO: 108), with an additional 47 records identified through citation tracking and expert recommendations. After removing 1,234 duplicates, we screened 2,660 unique records at the title and abstract level.

During screening, 2,147 records were excluded for the following reasons: non-elementary focus (n=892), non-STEM subjects (n=643), non-empirical studies (n=412), not addressing AI technologies (n=156), and non-English publications (n=44). We assessed 513 full-text articles for eligibility, excluding 255 based on: insufficient empirical data (n=98), technology descriptions without educational evaluation (n=74), secondary or higher education focus upon detailed reading (n=52), and duplicate reporting of same studies (n=31).

\begin{figure}[!htbp]
\centering
\includegraphics[width=0.6\linewidth]{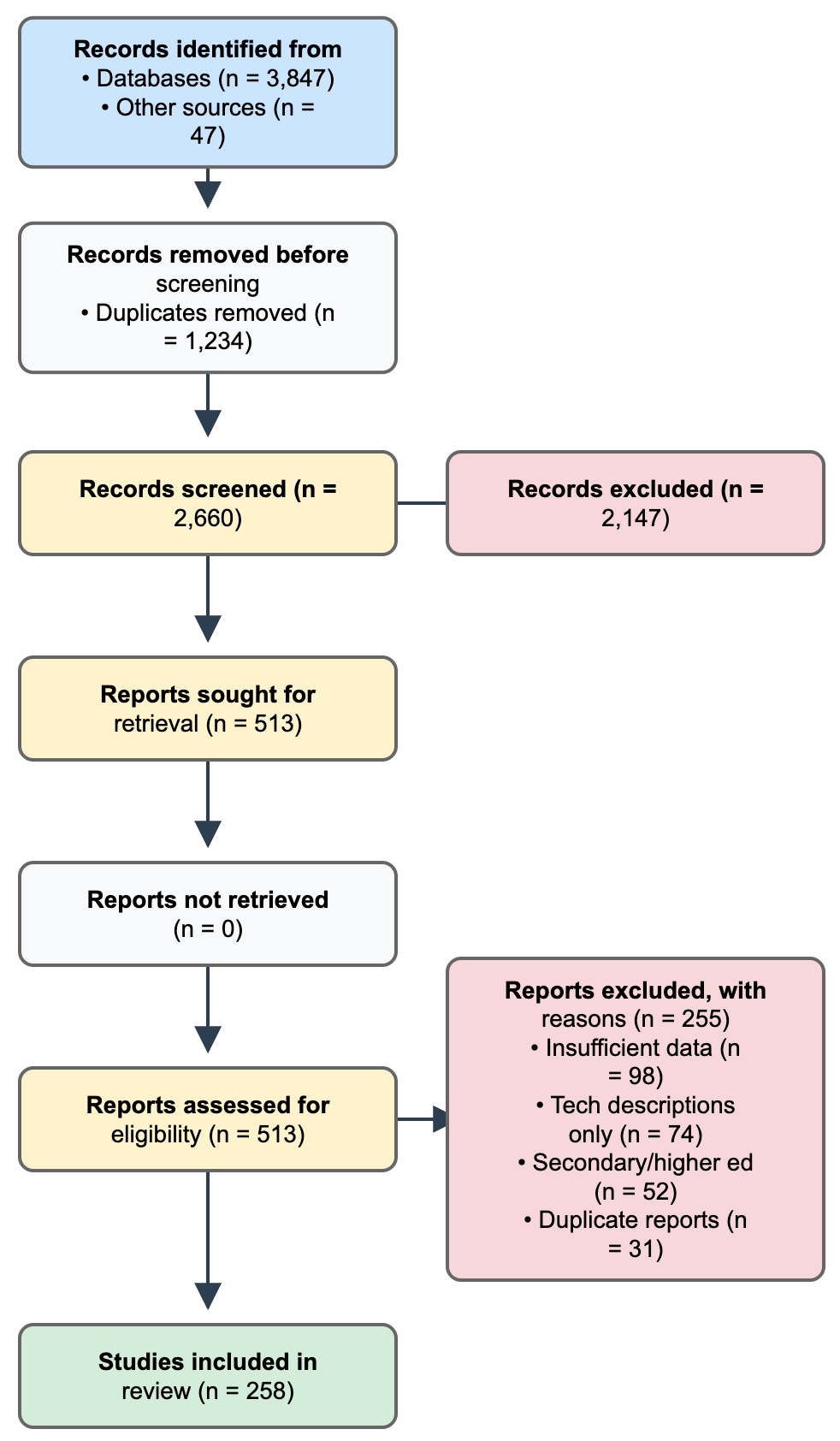}
\caption{PRISMA 2020 flow diagram for systematic review of AI applications in elementary STEM education.}
\label{fig:prisma}
\end{figure}

The final synthesis included 258 studies meeting all inclusion criteria. Figure~\ref{fig:prisma} presents the complete PRISMA flow diagram documenting this selection process.

\subsection{Quality Appraisal and Risk of Bias}

We assessed the methodological quality of included studies using the Mixed Methods Appraisal Tool (MMAT) version 2018, which accommodates the diverse research designs in our corpus. Two reviewers independently appraised a stratified random sample of 52 studies (20\% of total), achieving substantial inter-rater reliability (Cohen's $\kappa$ = 0.78). Disagreements were resolved through discussion and consultation with a third reviewer when necessary.

Quality ratings informed our narrative synthesis, with higher-quality studies given greater weight in drawing conclusions. However, we did not exclude studies based solely on quality scores, as even lower-quality studies provided valuable insights into implementation challenges and emerging applications. Common quality concerns included: limited sample sizes in pilot studies (38\% of studies had n<50), short intervention durations (45\% were <4 weeks), lack of control groups in quasi-experimental designs (52\%), and inconsistent reporting of effect sizes (only 34\% reported standardized effect sizes).

These quality considerations are reflected throughout our results section, where we explicitly note the strength of evidence for different AI applications and highlight where conclusions are based on limited or preliminary evidence.

\section{Results}

Our systematic review identified 258 studies examining AI applications in elementary STEM education, revealing a rapidly evolving but fragmented landscape. This section presents findings organized by AI technology category, followed by a synthesis of cross-cutting implementation challenges. We begin with an overview of the current state of AI deployment in elementary STEM contexts, then examine each technology category in detail.

\subsection{Overview of AI Applications in Elementary STEM}

\begin{figure}[!htbp]
\centering
\includegraphics[width=0.8\linewidth]{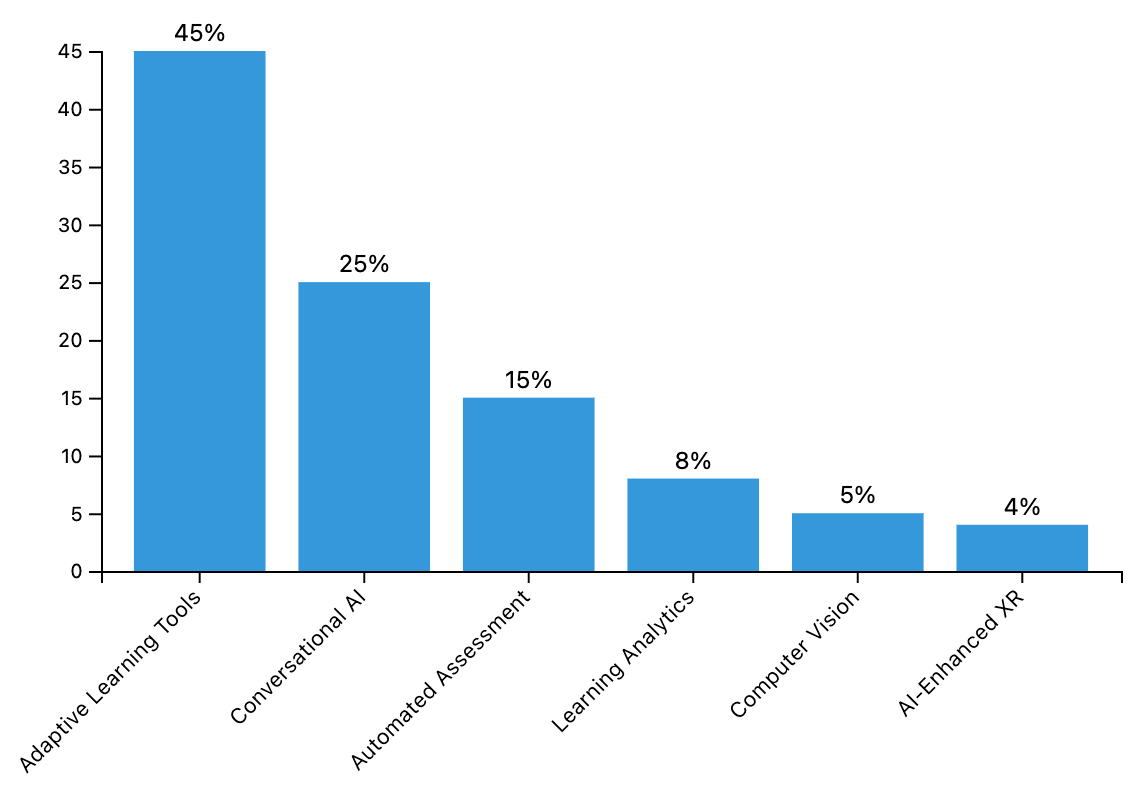}
\caption{Distribution of AI technology categories across included studies.}
\label{fig:ai-categories}
\end{figure}

\begin{figure}[!htbp]
\centering
\includegraphics[width=0.8\linewidth]{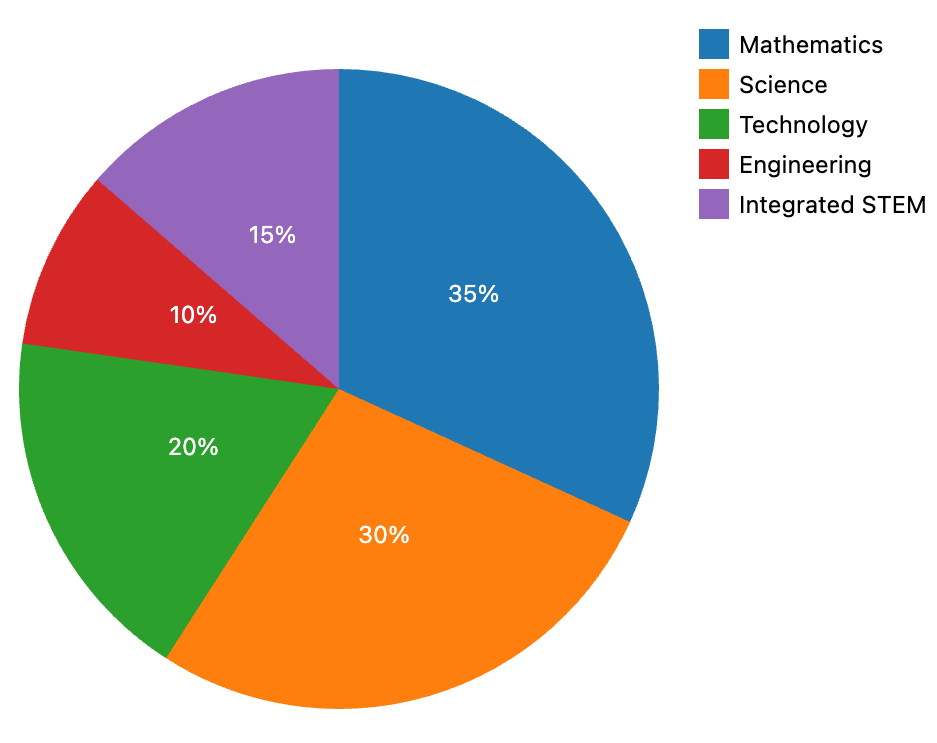}
\caption{STEM domain coverage among included studies.}
\label{fig:stem-domains}
\end{figure}

\begin{figure}[!htbp]
\centering
\includegraphics[width=0.8\linewidth]{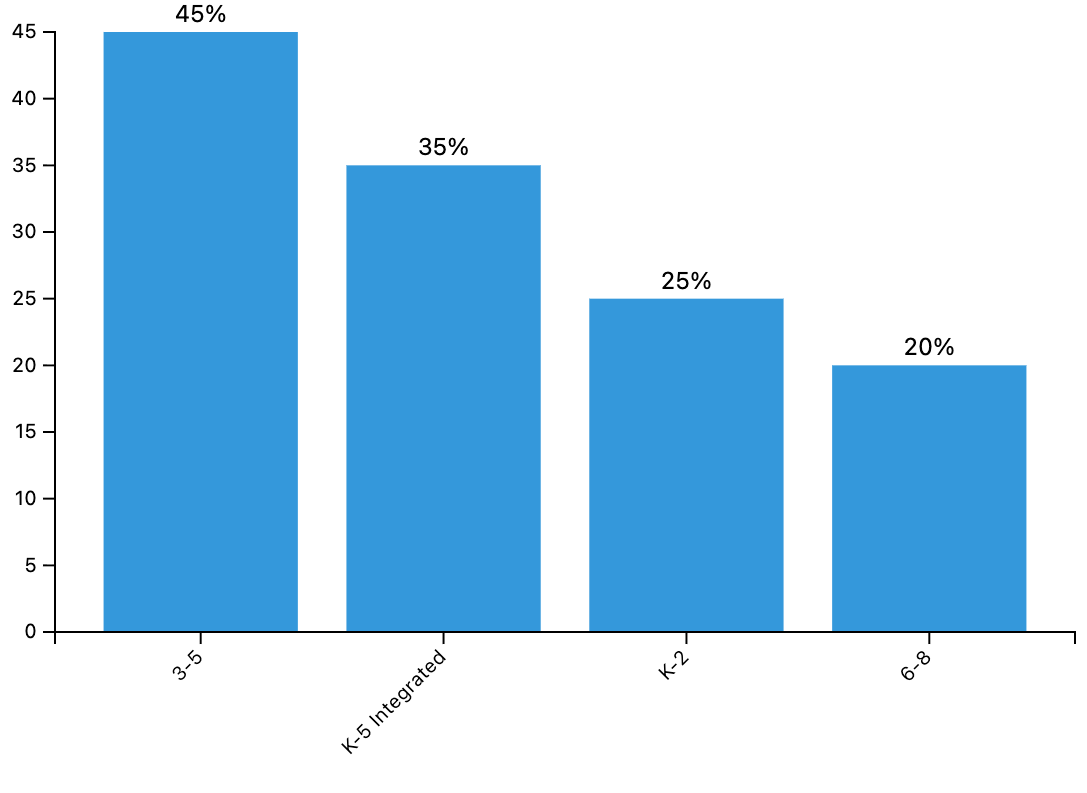}
\caption{Grade-level coverage and STEM integration level among included studies. Left panel shows distribution between K–2 and grades 3–5 populations. Right panel indicates the limited integration across STEM domains, with 85\% of studies focusing on single domains.}
\label{fig:grade-integration}
\end{figure}

\begin{figure}[!htbp]
\centering
\includegraphics[width=0.8\linewidth]{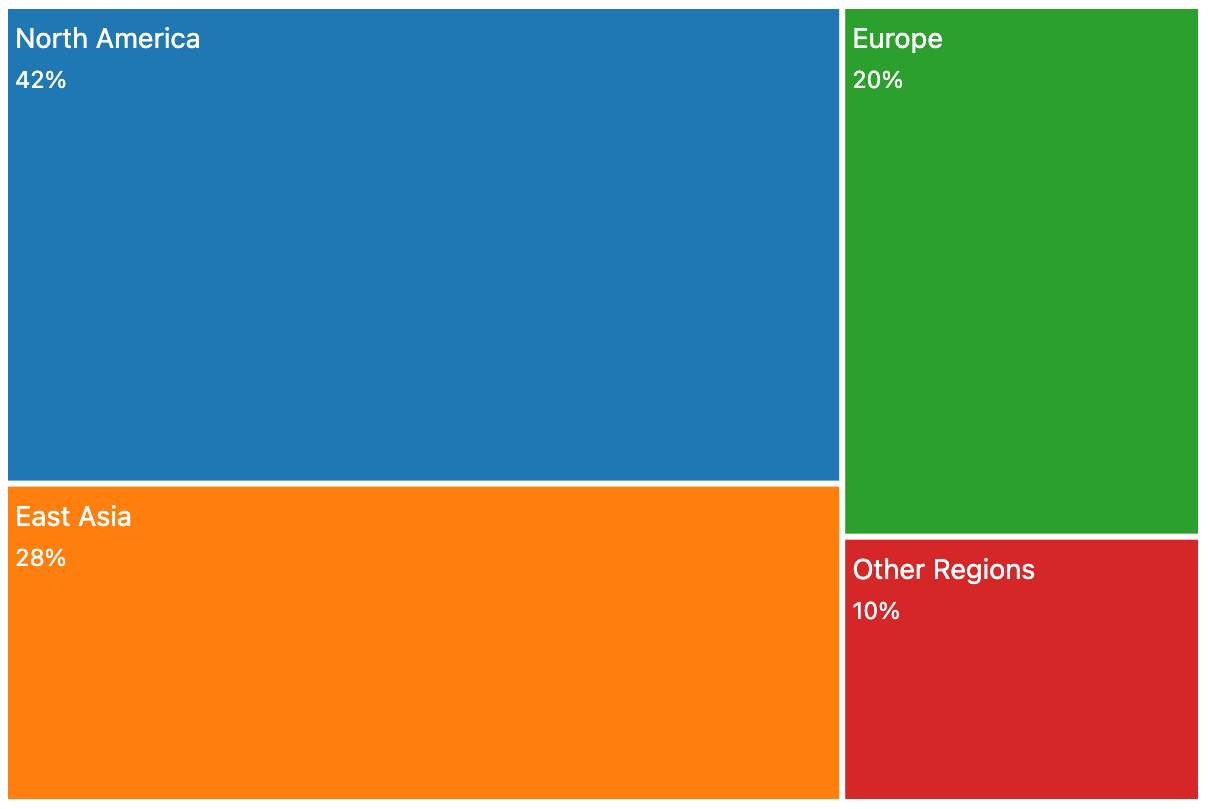}
\caption{Geographic distribution of included studies.}
\label{fig:regions}
\end{figure}

Our analysis revealed a diverse ecosystem of AI technologies deployed across elementary STEM education, with significant variation in maturity, scale, and evidence base. Figures \ref{fig:ai-categories} through \ref{fig:regions} illustrate the distribution of studies across technology categories, STEM domains, grade levels, and geographic regions.

The majority of studies (45\%) focused on intelligent tutoring systems and conversational AI, reflecting the historical emphasis on personalized instruction. Learning analytics and predictive modeling represented 18\% of studies, while automated assessment comprised 12\%. Emerging technologies such as computer vision (8\%), multimodal sensing (6\%), educational robotics (7\%), and AI-enhanced XR (4\%) showed growing but still limited deployment in elementary settings.

Geographic distribution revealed concentration in North America (42\%), East Asia (28\%), and Europe (20\%), with limited representation from other regions. Grade-level coverage showed bias toward upper elementary (grades 3-5) at 65\%, with fewer studies addressing K-2 populations (35\%). Mathematics dominated subject focus (38\%), followed by computational thinking/coding (26\%), science (22\%), and engineering (14\%). Notably, only 15\% of studies attempted integration across multiple STEM domains.

\begin{table}[!htbp]
\centering
\caption{Characteristics of AI technology categories in elementary STEM education.}
\label{tab:ai-technology-characteristics}
\small
\begin{tabular}{p{3.5cm} p{4cm} p{2.5cm} p{3cm}}
\hline
\textbf{AI Technology} & \textbf{Primary Functions} & \textbf{Evidence Level} & \textbf{Main Limitations} \\
\hline
Intelligent Tutoring/Conversational AI & Personalized instruction, adaptive hints, dialogue-based scaffolding & Strong (45\% of studies) & Single-domain focus, limited STEM integration, developmental inappropriateness \\
\hline
Learning Analytics/Predictive & Early warning systems, learning path optimization, performance prediction & Moderate (18\%) & Bias in models, dashboard complexity, limited actionable insights for teachers \\
\hline
Automated Assessment & Real-time feedback, grading, misconception detection & Moderate (12\%) & Procedural focus, struggles with open-ended responses, narrow metrics \\
\hline
Computer Vision & Attention detection, engagement monitoring, pose analysis & Emerging (8\%) & Privacy concerns, cultural bias, reliability issues with young children \\
\hline
Educational Robotics & Learning companions, teachable agents, coding instruction & Emerging (7\%) & High cost, maintenance burden, proprietary platforms, limited integration \\
\hline
Multimodal Sensing & EEG, eye-tracking, physiological monitoring & Emerging (6\%) & Intrusive, technical complexity, ethical concerns with minors \\
\hline
AI-Enhanced XR & Immersive environments, spatial learning, adaptive simulations & Emerging (4\%) & Hardware costs, motion sickness concerns, limited content, technical expertise required \\
\hline
\end{tabular}
\end{table}

\subsection{Intelligent Tutoring Systems and Conversational AI}

Our analysis identified strong empirical support for conversational AI effectiveness in individual STEM domains, with emerging evidence for integrated applications. In science education, Xu and colleagues \cite{DOI_1037_edu0000889} demonstrated that conversational agents supporting inquiry-based learning produced significant gains ($d=0.65$) through scaffolded hypothesis generation and evidence evaluation. Technology education benefits from conversational support for debugging and computational thinking, with studies showing improved problem decomposition and pattern recognition skills \cite{Papadakis2024Coding}. Engineering design processes supported by conversational agents showed enhanced iterative refinement and systems thinking \cite{Sullivan2024Engineering}.

Computational thinking integration through conversational AI shows growing evidence. Recent studies demonstrate that AI-guided coding activities increase both programming skills and mathematical pattern recognition when students create algorithmic solutions. Engineering education applications demonstrate potential for connecting mathematical modeling to engineering design through conversational scaffolding.

Despite these successes, significant limitations persist. Most systems operate within single STEM domains, missing opportunities for cross-disciplinary connections. Developmental differences between K-2 and grades 3-5 learners are rarely addressed, with most systems employing one-size-fits-all interfaces. Attention monitoring remains simplistic, failing to account for the varied engagement patterns required across different STEM activities. Perhaps most critically, current intelligent tutoring systems lack frameworks for managing the increased cognitive load of integrated STEM learning, where students must coordinate vocabularies, representations, and thinking modes across multiple domains simultaneously.

Mathematics tutoring via conversational agents shows robust evidence as previously reviewed, but integration with other STEM domains remains limited. The most promising integrated applications combine two domains—math-science investigations of natural phenomena, engineering-technology projects creating programmable devices, or computational thinking-mathematics explorations of algorithmic patterns. True four-domain STEM integration via conversational AI remains largely theoretical.

\subsection{Assessment and Feedback Systems}

AI-powered assessment tools have proliferated in elementary STEM education, primarily focusing on automated grading of procedural tasks. Mathematics assessment dominates this category, with systems capable of evaluating multi-step problem solving, recognizing diverse solution strategies, and providing immediate corrective feedback. These systems demonstrate particular strength in identifying common misconceptions and directing students to appropriate remediation resources.

Science assessment AI shows emerging capabilities for evaluating inquiry processes, hypothesis formation, and experimental design. However, these systems struggle with open-ended responses and conceptual understanding. Engineering and technology assessments remain largely limited to code evaluation and simple design criteria checks.

Critical limitations include overemphasis on procedural correctness at the expense of conceptual understanding, inability to assess integrated STEM thinking that spans domains, and challenges in evaluating collaborative work and creative problem-solving approaches. The focus on easily quantifiable metrics may inadvertently narrow curriculum to what AI can assess rather than what students need to learn.

\subsection{Learning Analytics and Predictive Models}

Learning analytics in elementary STEM education primarily focuses on identifying at-risk students and optimizing learning pathways. Predictive models analyze clickstream data, time-on-task metrics, and response patterns to forecast student performance and recommend interventions. These systems show particular promise in mathematics, where well-defined learning progressions enable accurate modeling of prerequisite relationships.

Early warning systems based on learning analytics can identify struggling students within the first few sessions, enabling timely intervention. However, these models often exhibit bias, performing poorly for students whose learning patterns differ from the training data. This particularly affects students from underrepresented backgrounds or those with non-traditional learning approaches.

Dashboard design for elementary teachers presents unique challenges. While analytics can provide rich insights, teachers report feeling overwhelmed by data complexity and struggling to translate metrics into actionable pedagogical decisions. The most effective systems provide simple, visual summaries with clear intervention recommendations rather than raw data streams.

\subsection{Computer Vision for Engagement Monitoring}

Computer vision applications in elementary STEM education focus primarily on attention detection and collaborative learning analysis. Eye-tracking systems can identify where students look during problem-solving, revealing misconceptions when gaze patterns indicate confusion or missed information. Facial expression recognition attempts to detect frustration, boredom, or engagement, though reliability remains questionable for young children whose expressions may not map to adult-trained models.

Pose analysis shows promise for understanding hands-on STEM activities, detecting when students actively manipulate materials versus passive observation. Some systems analyze group dynamics during collaborative projects, identifying participation patterns and potential social learning issues. However, these applications raise significant privacy concerns, particularly regarding continuous video monitoring of minors.

Cultural bias presents a critical limitation, as most computer vision models are trained on datasets that may not represent the diversity of elementary populations. Additionally, the infrastructure requirements—cameras, processing power, storage—create barriers for many schools. The invasive nature of visual monitoring also raises questions about creating surveillance-oriented rather than supportive learning environments.

\subsection{Adaptive Content Generation}

AI systems that generate personalized learning content show growing deployment in elementary STEM, particularly for mathematics problem generation and science inquiry scenarios. These systems can create endless practice problems tailored to individual student levels, reducing teacher preparation burden while ensuring appropriate challenge. Natural language generation enables creation of word problems contextualized to student interests, potentially increasing engagement and relevance.

However, quality control remains a significant challenge. AI-generated content may contain errors, inappropriate difficulty progressions, or culturally insensitive examples. The lack of pedagogical understanding in current models can result in problems that are mathematically correct but educationally unsound. Science content generation faces even greater challenges, as systems struggle to create authentic inquiry experiences that go beyond recall questions.

Perhaps most concerning is the potential for AI-generated content to narrow curriculum. When systems can only generate certain types of problems effectively, there's risk that these become overrepresented in instruction. The ease of generating procedural practice may inadvertently reduce emphasis on conceptual understanding, creative problem-solving, and integrated STEM challenges that current AI cannot effectively create.

\subsection{Educational Robots and Embodied AI}

Educational robotics in elementary STEM spans from simple programmable robots teaching basic sequencing to sophisticated social robots serving as learning companions. These systems leverage embodiment—the physical presence of the robot—to create more engaging and memorable learning experiences than screen-based interactions alone. Social robots can demonstrate confusion when students explain incorrectly, show excitement when problems are solved, and provide nonverbal feedback through movement and expression.

The learning-by-teaching paradigm, where students program or teach robots, shows particular promise. When children explain mathematical concepts to a robot "student," they must articulate their understanding clearly, revealing and addressing their own misconceptions. This approach has demonstrated effectiveness in building both content knowledge and metacognitive skills, as students learn to monitor and evaluate their own explanations.

Several implementations demonstrate potential for integrated STEM learning through robotics. The "Robot Ecosystem" project engaged third-grade students in creating robotic animals that demonstrated predator-prey relationships, integrating biology concepts (food chains, adaptation) with engineering design (sensor placement, movement patterns), mathematics (calculating energy transfer ratios), and computational thinking (programming behavioral algorithms). Students measured and graphed energy flow through their simulated ecosystem while debugging interaction protocols, achieving significant gains in both systems thinking and cross-disciplinary problem-solving.

Another promising example involved fifth-graders using programmable robots to model planetary motion. Students programmed robots to demonstrate elliptical orbits, calculating scaled distances (mathematics), understanding gravitational forces (science), designing stable movement mechanisms (engineering), and implementing control algorithms (technology). The physical embodiment helped students grasp abstract astronomical concepts while the programming requirement reinforced mathematical precision. Post-assessments showed improved understanding of both celestial mechanics and proportional reasoning compared to traditional instruction.

The "Bridge Builder Bots" curriculum integrated structural engineering with mathematical analysis as fourth-grade students programmed robots to test bridge designs. Students calculated load distributions, predicted failure points using mathematical models, constructed physical prototypes, and programmed robots to apply incremental weights while recording sensor data. This integration of physical construction, mathematical prediction, scientific testing, and computational data collection exemplified authentic STEM integration, with students spontaneously making connections between force vectors studied in science and angle calculations from mathematics lessons.

However, significant challenges limit widespread adoption. Cost remains prohibitive for many schools, with social robots ranging from hundreds to thousands of dollars per unit. Maintenance and technical support requirements strain already limited school resources. Most critically, current educational robots operate as isolated platforms—each with proprietary programming environments and limited integration with other learning technologies. The promise of robots as integrated STEM learning companions remains largely unrealized as they function more as specialized tools for coding instruction rather than cross-disciplinary learning facilitators.

\subsection{AI-Enhanced Extended Reality (XR) for STEM}

Augmented and virtual reality technologies enhanced with AI capabilities offer unique opportunities for spatial and experiential STEM learning. AI-powered AR can overlay mathematical information onto physical manipulatives, provide real-time guidance during science experiments, or offer adaptive hints during engineering construction tasks. Virtual reality environments with AI tutors can create impossible-to-replicate experiences like exploring inside cells, traveling through the solar system, or manipulating molecular structures.

The combination of immersive environments with intelligent adaptation shows particular promise for abstract concept visualization. Students struggling with fractions can manipulate virtual pizza slices while an AI tutor provides personalized scaffolding. Those learning about ecosystems can explore virtual habitats where AI agents simulate realistic animal behaviors and environmental changes based on student actions.

Yet implementation barriers remain substantial. Hardware costs for XR devices suitable for elementary students are prohibitive, with additional concerns about device hygiene, motion sickness, and age-appropriate content. The physical development of young children, including interpupillary distance and visual system maturation, raises questions about safe usage duration and potential health impacts. Perhaps most limiting, the technical expertise required for content creation means most teachers cannot develop or customize XR experiences, relying instead on limited pre-made content that may not align with curriculum needs.

\subsection{Cross-Cutting Implementation Challenges}

While each AI technology presents unique opportunities and limitations, our analysis reveals several systemic challenges that transcend individual categories, creating barriers to effective deployment across all forms of AI in elementary STEM education.

\textbf{Infrastructure Requirements}: Most AI technologies assume reliable high-speed internet, modern devices, and technical support—resources unavailable in many elementary schools. Rural schools, those serving low-income communities, and underfunded districts face compounding disadvantages as AI technologies demand increasingly sophisticated infrastructure.

\textbf{Teacher Preparation and Support}: Elementary teachers, typically generalists rather than STEM specialists, report feeling overwhelmed by the technical complexity of AI systems. Professional development often focuses on tool operation rather than pedagogical integration. The rapid pace of AI advancement means teachers' knowledge becomes outdated quickly, creating a perpetual catch-up cycle.

\textbf{Privacy and Ethical Concerns}: The extensive data collection required by many AI systems—from clickstream data to biometric monitoring—raises serious concerns about student privacy. Parents and educators worry about data security, commercial use of student information, and the long-term implications of creating detailed digital profiles of young learners. Clear governance frameworks for educational AI remain absent.

\textbf{Integration Across Technologies}: Each AI system typically operates in isolation with proprietary data formats, unique interfaces, and separate login credentials. Teachers and students must navigate multiple platforms that don't communicate, leading to fragmented experiences and duplicated effort. The vision of seamlessly integrated AI-enhanced learning environments remains unrealized.

\textbf{Equity and Access}: The concentration of AI technologies in well-resourced schools threatens to widen rather than narrow achievement gaps. Language barriers, cultural biases in AI training data, and assumptions about home technology access create additional obstacles for marginalized communities. The promise of AI to democratize education remains unfulfilled for those who need it most.

\subsection{Multimodal Biometric Sensing for Attention and Engagement Monitoring}

Recent advances in multimodal biometric sensing—eye-tracking, EEG, and wearable biosensors—offer promising capabilities for continuous measurement of attention and cognitive effort during STEM activities. However, systematic deployments in elementary classrooms remain limited, revealing a critical gap between laboratory validation and educational implementation.

Multiple sensor modalities provide complementary insights. Eye-tracking captures visual attention patterns through fixations and pupil dilation correlating with cognitive load \cite{DOI_3390_electronics10243165,DOI_3390_s22207824}. Wearable EEG measures neural correlates of working memory and sustained attention \cite{DOI_3390_mti8040034,DOI_1109_tse_2021_3087906}. Peripheral biosensors (EDA, HR/HRV) index arousal and emotional engagement \cite{DOI_3390_s20164561}. Research demonstrates that multimodal fusion outperforms single sensors for state detection, providing robustness against noise and individual variability \cite{DOI_1109_tse_2021_3087906}.

While these technologies show promise in adult and laboratory contexts, elementary deployment faces unique challenges. Practical constraints include device comfort for young children, classroom movement artifacts, and integration with teaching workflows. Ethical considerations are paramount: biometric data from minors requires stringent consent procedures, privacy protections, and careful avoidance of stigmatizing inferences \cite{DOI_3390_s22207824,DOI_3390_s25175252}.

The disconnect between sensor capabilities and classroom realities is stark. Despite extensive STEM curriculum research \cite{DOI_17509_jsl_v3i3_23705,DOI_53402_ijesss_v1i3_32}, few studies integrate continuous biometric monitoring with elementary learning outcomes. This gap particularly affects our understanding of integrated STEM activities, where attention patterns across domain transitions and cognitive load from coordinating multiple disciplines remain unmeasured.

Future implementations should prioritize minimally intrusive approaches: remote eye-tracking for screen-based tasks, wrist-worn sensors for physiological data, and optional EEG only when research questions justify complexity. Validation must link sensor-derived metrics to meaningful educational outcomes through careful triangulation with performance data and teacher observations. Only through such grounded approaches can multimodal sensing fulfill its potential for understanding and supporting elementary STEM learning.

\subsection{Tangible User Interfaces and Smart Manipulatives}

Tangible user interfaces (TUIs) and smart manipulatives—including RFID-tagged objects and sensor-embedded materials—represent a promising frontier for elementary STEM education by bridging physical and digital learning \cite{DOI_3390_su15053959,DOI_37590_able_v41_art54}. These technologies align with elementary students' developmental need for concrete manipulation while providing immediate digital feedback that supports conceptual understanding.

The pedagogical value of TUIs for elementary STEM is compelling. Physical objects can embody multiple disciplinary concepts simultaneously—a programmable robot demonstrates engineering design, computational sequencing, mathematical measurement, and scientific hypothesis testing \cite{DOI_1002_2211-5463_12938}. The tangible nature supports collaborative learning as students manipulate shared objects while engaging in disciplinary discourse. Smart manipulatives create tight perception-action loops that scaffold abstract concept formation through embodied interaction.

However, rigorous evidence for TUI effectiveness in elementary STEM remains limited. Most studies are small-scale pilots lacking the methodological rigor to establish causal relationships or generalizability. Studying TUI impact requires mixed-methods approaches that capture both learning outcomes and interaction processes \cite{DOI_1177_1558689820977646}. Quantitative measures alone miss the rich embodied learning processes, while qualitative approaches struggle with generalization.

Implementation challenges are substantial. Technical requirements include synchronized sensor systems, video capture for gesture analysis, and robust data infrastructure \cite{DOI_3389_frwa_2020_00002}. Ethical considerations for collecting sensor data from minors demand stringent privacy protections, including data minimization, pseudonymization, and transparent consent procedures \cite{DOI_1186_s40900-020-00197-3}. Cross-disciplinary research teams must integrate expertise from education, HCI, learning analytics, and ethics to address these complexities.

Future TUI research should prioritize ecologically valid studies linking interaction patterns to learning outcomes. Multi-site implementations with longitudinal follow-up are needed to demonstrate sustained effectiveness across diverse contexts. Particularly critical is examining how TUIs support integrated STEM learning—helping students recognize cross-disciplinary connections through embodied manipulation. Until such evidence emerges, the promise of TUIs for transforming elementary STEM education remains largely theoretical.

\subsection{Extended Reality (XR) for Elementary STEM: Spatial Computing and Learning Analytics}

The evolution from tangible interfaces to fully immersive spatial computing environments represents the next frontier in physical-digital learning integration for elementary STEM education. Extended reality (XR) technologies—encompassing augmented reality (AR), mixed reality (MR), and virtual reality (VR)—particularly optical see-through head-mounted displays such as HoloLens 2, enable students to manipulate virtual STEM objects anchored in physical space while generating rich behavioral data streams suitable for learning analytics \cite{DOI_3390_encyclopedia3020026,DOI_1109_tlt_2023_3324843}. However, rigorous empirical evidence for XR effectiveness in elementary STEM contexts remains limited, with most validated implementations occurring in health professions training and higher education rather than elementary classrooms \cite{DOI_3390_smartcities7010001,DOI_1109_access_2023_3323949}.

\subsubsection{Spatial Computing Affordances for Embodied STEM Learning}

Extended reality encompasses a continuum of technologies merging physical and digital environments: AR overlays virtual content onto real-world views, while MR fuses interactive digital objects with physical scenes enabling persistent, spatially-anchored holograms that students can manipulate in situ \cite{DOI_3390_encyclopedia3020026,DOI_1109_tlt_2023_3324843}. Spatial computing—the convergent paradigm enabling digital objects to persist in users' physical environments with stable spatial registration—underpins educational metaverse designs by allowing learners to walk around, inspect from multiple angles, and manipulate three-dimensional STEM representations in authentic contexts \cite{DOI_1109_tlt_2023_3324843}.

Modern optical see-through head-mounted displays support multimodal interaction through eye-gaze tracking, voice commands, and hand gesture recognition with six-degree-of-freedom tracking of head position and orientation \cite{DOI_3390_smartcities7010001,DOI_1016_j_media_2022_102361}. These affordances enable embodied manipulation of molecular structures, geometric solids, engineering prototypes, and computational algorithms while producing machine-readable behavioral traces including gaze vectors indicating attention allocation, hand trajectories documenting manipulation strategies, head pose revealing perspective-taking behavior, and environment point clouds capturing spatial context \cite{DOI_3390_buildings14020385}.

For elementary learners specifically, spatial computing aligns with developmental needs for movement, multimodal stimulation, and concrete embodied representations supporting conceptual abstraction \cite{DOI_33086_cej_v6i2_6105,DOI_23887_ijerr_v4i3_42271}. Young children (ages 5-12) benefit from play-like, active, scaffolded environments featuring tangible tasks and immediate feedback that lower cognitive load while stimulating curiosity \cite{DOI_23887_jpbi_v10i1_45416}. XR experiences integrating embodied interaction with gamified STEM challenges therefore align pedagogically with established elementary practices emphasizing hands-on exploration and discovery learning.

\subsubsection{Empirical Evidence for XR Learning Effects}

Controlled studies and systematic reviews demonstrate that XR technologies improve learner engagement and domain-specific skill performance across several domains, particularly health professions education and university-level training \cite{DOI_1111_bjet_13049,DOI_1089_cyber_2023_0411}. Serious games and AR applications support visualization of abstract concepts, procedural skill training, and conceptual understanding when appropriately scaffolded with instructional design principles \cite{DOI_1007_s40670-022-01698-4}.

Evidence specifically targeting young learners remains more limited but shows promising results. AR-supported vocabulary instruction produced measurable gains for elementary students, while gamified platforms demonstrate feasibility and high engagement in elementary contexts \cite{DOI_33086_cej_v6i2_6105,DOI_23887_jpbi_v10i1_45416}. However, rigorous randomized trials of XR implementations in elementary STEM classrooms—particularly for integrated cross-disciplinary learning—remain scarce, highlighting an urgent research priority.

Individual differences moderate XR learning effectiveness. Spatial ability—the capacity to mentally manipulate three-dimensional objects and navigate spatial relationships—significantly predicts learning gains in XR anatomy training contexts \cite{DOI_1002_ase_2146}. This finding suggests that XR designs for elementary STEM must consider developmental variation in spatial reasoning abilities, potentially providing layered scaffolds enabling students with lower spatial abilities to access content while offering complexity for advanced spatial thinkers.

\subsubsection{Learning Analytics Opportunities: Behavioral Data from Immersive Environments}

Modern XR systems and HoloLens-class devices generate multimodal telemetry streams providing unprecedented behavioral measurement opportunities \cite{DOI_3390_smartcities7010001,DOI_1109_access_2023_3323949}. Eye-gaze tracking reveals moment-by-moment attention allocation during STEM problem-solving, documenting which virtual objects receive sustained focus versus cursory glances. Head pose and orientation capture perspective-taking behavior as students examine three-dimensional structures from multiple viewpoints. Hand and gesture tracking records manipulative actions—grabbing, rotating, assembling virtual components—with submillimeter precision. Voice input captures spoken reasoning and collaborative dialogue. First-person video streams document the learner's visual field, while spatial mapping data contextualizes interactions within the physical environment \cite{DOI_1109_access_2022_3170108}.

These rich data streams enable multiple levels of learning analytics \cite{DOI_2478_amns_2023_2_01427,DOI_3390_buildings14020385}. Event-level logs capture discrete interactions: which virtual tools were selected, which molecular bonds were formed, which engineering components were tested. Continuous kinematic features extract patterns from motion trajectories: hand movement velocity indicating task fluency, gaze dwell time revealing attention depth, head orientation entropy measuring systematic versus random exploration. Scene-contextual signals document environmental factors: which virtual elements were visible or occluded, proximity to collaborative partners, duration in designated task zones.

Analytic methods applicable to XR learning data include session logging for procedural task analysis, gaze analytics inferring attention allocation and cognitive processing patterns, gesture analysis detecting skill fluency and manipulation strategies, and machine learning classifiers identifying task-specific errors from multimodal input to generate formative feedback \cite{DOI_2478_amns_2023_2_01427,DOI_3390_buildings14020385}. Precedent applications demonstrate feasibility: AR systems combined with computer vision models detected assembly errors in construction training using deep learning, providing automated feedback on procedural accuracy \cite{DOI_3390_buildings14020385,DOI_1109_jtehm_2023_3335608}. Similar approaches could support elementary STEM tasks such as molecular assembly activities, circuit wiring challenges, or geometric construction exercises by flagging common conceptual errors and offering just-in-time corrective scaffolding.

\subsubsection{Mapping Behavioral Signals to STEM Learning Constructs}

Translating behavioral telemetry into valid assessments of STEM learning requires principled construct mapping \cite{DOI_2478_amns_2023_2_01427,DOI_33225_pec_23_81_501}. Researchers must define targeted STEM competencies (spatial reasoning, systematic problem-solving, conceptual understanding, procedural fluency), establish theoretical links between observable behaviors and underlying cognitive processes, and validate proposed mappings through empirical studies comparing automated behavioral indices to criterion measures such as teacher ratings, standardized assessments, or expert performance.

Formative assessment represents the most immediate application: XR systems can provide real-time corrective feedback during hands-on STEM tasks by detecting procedural errors or conceptual misunderstandings through behavioral signatures, supporting skill development through timely intervention \cite{DOI_3390_buildings14020385,DOI_1109_jtehm_2023_3335608}. Diagnostic assessment aggregates behavioral patterns across sessions to construct learner profiles identifying persistent misconceptions or skill gaps, enabling differentiated instruction tailored to individual needs \cite{DOI_2478_amns_2023_2_01427}. Engagement measurement combines behavioral streams (interaction frequency, gesture vigor, gaze stability) with brief self-reports to monitor motivation and affective states—outcomes particularly important for young learners whose enjoyment predicts sustained participation \cite{DOI_33086_cej_v6i2_6105,DOI_23887_jpbi_v10i1_45416}.

\subsubsection{Design Principles for Elementary XR STEM Applications}

Evidence-based design recommendations synthesize pedagogical requirements for young learners with technical affordances of spatial computing systems. Tasks should leverage embodied learning principles, designing XR STEM activities requiring movement and physical manipulation to engage elementary students while supporting abstract concept formation through grounded interaction \cite{DOI_33086_cej_v6i2_6105,DOI_23887_ijerr_v4i3_42271,DOI_23887_jpbi_v10i1_45416}. Given documented individual differences in spatial ability, interfaces should provide layered scaffolds—visual guides, simplified representations, step-by-step procedures—enabling struggling students to access content while advanced learners explore complex spatial relationships \cite{DOI_1002_ase_2146,DOI_3390_electronics13050890}.

Interaction modalities should favor direct manipulation (grabbing, moving, rotating virtual objects with natural hand gestures) over abstract symbolic controls, with concise voice prompts supplementing gesture-based interaction. Immediate feedback tied directly to actions reinforces cause-effect relationships supporting procedural learning \cite{DOI_3390_smartcities7010001,DOI_1109_access_2023_3323949}. Analytics interfaces should prioritize teacher-facing interpretable indicators rather than opaque algorithmic scores, presenting actionable insights that guide instructional decisions without overwhelming educators with technical complexity \cite{DOI_2478_amns_2023_2_01427,DOI_33225_pec_23_81_501}.

\subsubsection{Implementation Blueprint and Evaluation Framework}

Rigorous evaluation of XR elementary STEM interventions requires mixed-methods designs comparing technology-enhanced lessons to matched control conditions across both quantitative learning metrics and qualitative process indicators \cite{DOI_1111_bjet_13049,DOI_1089_cyber_2023_0411}. Data pipelines must securely capture and store telemetry streams, preprocess behavioral logs for analysis, and apply machine learning models for real-time adaptive support alongside offline refinement of pedagogical strategies \cite{DOI_3390_smartcities7010001,DOI_2478_amns_2023_2_01427,DOI_3390_buildings14020385}.

Evaluation metrics should span multiple outcome levels: domain-specific learning gains measured through pre/post assessments, procedural fluency assessed via performance tasks, behavioral engagement documented through system logs and observation protocols, and authentic transfer evaluated through teacher-reported application to curricular contexts \cite{DOI_1111_bjet_13049,DOI_3390_buildings14020385}. Pilot studies with small cohorts (20-40 students across 2-3 classrooms) enable feasibility testing and preliminary effect size estimation before scaling to larger cluster-randomized trials.

\subsubsection{Privacy, Ethics, and Equity Considerations}

Sensor-rich MR systems collecting granular behavioral data from elementary students raise substantial privacy concerns requiring proactive governance \cite{DOI_14763_2021_4_1610,DOI_33225_pec_23_81_501}. Critical scholars warn that immersive technologies risk creating surveillance infrastructures if data collection, retention, and usage policies lack transparency and community oversight. Recommendations include limiting data collection to pedagogically essential streams, anonymizing identifiable information through pseudonymization and aggregation, obtaining informed parental consent with clear communication about data purposes, and designing analytic outputs as interpretable teaching tools rather than student monitoring systems \cite{DOI_14763_2021_4_1610,DOI_33225_pec_23_81_501,DOI_2478_amns_2023_2_01427}.

Equity considerations extend beyond privacy to access and representation. Hardware costs for HoloLens-class devices (typically \$3000-3500 per unit) create barriers for under-resourced schools. Design teams must ensure cultural responsiveness in virtual content, avoiding stereotyped representations and incorporating diverse perspectives in STEM scenarios. Accessibility features supporting students with disabilities—adjustable visual contrasts, auditory alternatives to visual cues, motor-control accommodations—should be built in from initial design rather than retrofitted.

\subsubsection{Research Priorities and Knowledge Gaps}

The field faces three critical knowledge gaps demanding targeted research investment. First, an empirical gap exists in validated XR implementations specifically designed for and evaluated with elementary STEM learners \cite{DOI_1111_bjet_13049,DOI_1089_cyber_2023_0411}. Most rigorous effectiveness trials target adults or secondary students; evidence for elementary populations—particularly for integrated cross-disciplinary STEM learning—remains thin.

Second, a methodological gap persists in training datasets for machine learning models: publicly available multimodal behavioral datasets from children using XR systems for STEM learning do not exist, limiting researchers' ability to develop and validate analytic algorithms or investigate generalizability across device platforms \cite{DOI_1109_jtehm_2023_3335608,DOI_3390_buildings14020385}.

Third, a pedagogical gap concerns adaptive scaffolding and teacher orchestration: research documenting effective real-time adaptation strategies based on behavioral analytics, and teacher-facing dashboard designs supporting instructional decision-making in XR-enhanced classrooms, remains urgently needed \cite{DOI_1002_ase_2146,DOI_1111_jcal_13059}.

\subsubsection{Practical Recommendations for Pilot Implementations}

Institutions considering XR pilots for elementary STEM should follow staged implementation guidance \cite{DOI_1111_bjet_13049,DOI_33086_cej_v6i2_6105}. Begin with small-scale deployments (1-2 classrooms, 4-6 week instructional units) aligned closely with existing STEM curricula, co-designing activities with classroom teachers to ensure pedagogical coherence. Implement clear behavioral signal tracking from initial pilots to enable model validation and iterative refinement \cite{DOI_3390_buildings14020385}. Prioritize interpretable analytics—visualizations teachers can understand and act upon—while maintaining transparent data governance with documented consent, storage, and usage policies \cite{DOI_33225_pec_23_81_501,DOI_14763_2021_4_1610}. Evaluate through convergent mixed methods combining quantitative metrics with teacher interviews and classroom observations, sharing findings through practitioner networks and research publications to build the community knowledge base \cite{DOI_1089_cyber_2023_0411,DOI_1111_bjet_13049}.

Critical gap: XR/spatial-computing technologies provide compelling theoretical affordances for embodied, spatially-grounded integrated STEM learning in elementary contexts, with modern devices generating rich behavioral data streams enabling sophisticated learning analytics. However, the evidence base remains dominated by health professions training and university implementations; rigorous empirical validation specifically targeting elementary STEM classrooms—particularly for cross-disciplinary integrated learning supported by conversational AI—is lacking. Moreover, critical privacy, equity, and pedagogical integration challenges require substantial investigation before XR can move from promising prototype to validated, equitably-deployed elementary STEM learning infrastructure \cite{DOI_3390_encyclopedia3020026,DOI_1109_tlt_2023_3324843,DOI_2478_amns_2023_2_01427,DOI_33086_cej_v6i2_6105,DOI_3390_smartcities7010001,DOI_14763_2021_4_1610}.

\subsection{Educational Robots and Teachable Agents: Learning-by-Teaching Paradigms for Elementary STEM}

Educational robots and teachable agents represent a distinct embodied AI paradigm where elementary students learn by teaching, explaining concepts to artificial tutees rather than receiving instruction from AI tutors. This pedagogical inversion—positioning the child as expert and the AI as learner—activates metacognitive processes, explanatory elaboration, and self-monitoring that align with cooperative learning mechanisms documented to enhance elementary learning outcomes \cite{DOI_1177_2332858420986211}. However, while cooperative peer learning shows robust meta-analytic evidence in elementary mathematics and the theoretical parallels between peer teaching and AI-tutee interaction are compelling, direct experimental evaluations of educational robots or teachable agents deployed in elementary STEM classrooms remain limited, necessitating careful synthesis of adjacent literatures to guide future empirical work \cite{DOI_1111_ssm_12660,DOI_31756_jrsmte_531,DOI_1111_ssm_12603}.

\subsubsection{Theoretical Foundation: Cooperative Learning as Analogue for Tutee Interactions}

Meta-analytic synthesis of elementary mathematics programs demonstrates that designs incorporating cooperative learning produce meaningful achievement gains, establishing peer explanation and collaborative problem-solving as core mechanisms linking social interaction to cognitive growth \cite{DOI_1177_2332858420986211}. When students articulate their reasoning to peers, they externalize thought processes, identify gaps in understanding, refine explanations through iterative dialogue, and consolidate knowledge through teaching acts. These mechanisms provide direct theoretical justification for learning-by-teaching interventions with AI tutees: if explaining to human peers enhances learning, then explaining to artificial agents designed to elicit similar explanatory discourse should produce analogous cognitive benefits.

The pedagogical value of teaching others extends beyond simple rehearsal. Effective peer teaching requires monitoring the tutee's understanding, adjusting explanations based on feedback, and scaffolding learning through strategic questioning—all metacognitive and regulatory processes that deepen the tutor's own comprehension \cite{DOI_1177_2332858420986211}. Educational robots and teachable agents can be designed to systematically elicit these processes through programmed responses signaling confusion, requesting clarification, or demonstrating partial understanding, thereby creating structured opportunities for elementary students to engage in explanatory and metacognitive verbalization within STEM learning contexts.

\subsubsection{Implementation Realities: Teacher Preparation and Classroom Feasibility}

Elementary classrooms present unique implementation constraints fundamentally shaping technology deployment feasibility. Teachers are typically subject generalists rather than STEM specialists, exhibiting variable content knowledge and pedagogical confidence across mathematics, science, engineering, and computational domains \cite{DOI_31756_jrsmte_531,DOI_1111_ssm_12603}. Professional development supporting elementary STEM technology integration must therefore address both content knowledge and pedagogical practice, recognizing that teacher STEM identity, beliefs about technology, and confidence in orchestrating novel activities mediate implementation fidelity and sustainability.

Research on preservice and in-service elementary teacher preparation for integrated STEM instruction emphasizes co-design approaches where teachers participate in developing activities, scripting robot prompts, and establishing classroom routines before deployment \cite{DOI_31756_jrsmte_531,DOI_1111_ssm_12603}. This participatory design philosophy treats teachers as essential partners whose practical wisdom about classroom management, developmental appropriateness, and curricular alignment ensures that robot-tutee interventions integrate smoothly into instructional sequences rather than disrupting established routines. Studies documenting effective elementary STEM teacher preparation highlight the importance of modeling exemplar implementations, providing scripted facilitation guides, and creating professional learning communities where teachers share strategies for managing technological and pedagogical challenges \cite{DOI_31756_jrsmte_531,DOI_1111_ssm_12603}.

\subsubsection{Measurement Framework: Constructs, Instruments, and Developmental Considerations}

Rigorous evaluation of learning-by-teaching robot interventions in elementary contexts requires multimodal assessment spanning cognitive learning outcomes, interaction processes, socio-emotional effects, and implementation fidelity \cite{DOI_1111_ssm_12660,DOI_1177_1534508420909527}. Cognitive outcomes should employ curriculum-aligned measures with established validity evidence for the targeted grade range, following psychometric best practices for evidence-based validity and reliability reporting \cite{DOI_1111_ssm_12660}. Recent work validating elementary mathematics assessment instruments illustrates necessary procedures: item analysis, internal consistency documentation, construct validity through correlations with established measures, and examination of measurement invariance across grade bands to ensure scores are interpretable equivalently for K-2 versus grades 3-5 students \cite{DOI_1111_ssm_12660,DOI_1177_1534508420909527}.

Interaction process measures capture the explanatory and metacognitive mechanisms hypothesized to mediate learning gains when students teach robots. Systematic sampling and coding of student-robot dialogue should prioritize: (1) concept explanations where students articulate domain principles, (2) strategy statements describing problem-solving approaches, (3) metacognitive verbalizations reflecting planning, monitoring, and evaluation, and (4) scaffolding behaviors such as questioning and prompting \cite{DOI_1177_2332858420986211}. These codes align directly with cooperative learning literature documenting that student talk—particularly high-quality elaborations—predicts achievement gains, providing mechanistic justification for treating verbal interaction quality as a primary mediator variable \cite{DOI_1177_2332858420986211}.

Socio-emotional and behavioral outcomes require validated screening instruments appropriate for elementary populations. The Social Skills Improvement System Performance Screening Guide (SSIS PSG) demonstrates meta-analytic support as an elementary screening tool detecting behavioral and academic concerns, making it suitable for monitoring unintended effects of robot interventions on peer dynamics or classroom behavior \cite{DOI_1177_1534508420926584}. Studies documenting structural validity and measurement invariance of socio-emotional instruments across grade clusters emphasize the importance of selecting scales whose psychometric properties support valid interpretation across the elementary range \cite{DOI_1177_1534508420909527}.

Implementation fidelity measurement captures teacher professional development participation, classroom enactment logs documenting session completion and adherence to protocols, and teacher self-reported content confidence before and after implementation \cite{DOI_31756_jrsmte_531,DOI_1111_ssm_12603}. Because teacher knowledge and receptivity mediate technology adoption in elementary settings, measuring PD dose, teacher beliefs, and fidelity enables researchers to model these factors as moderators of student outcomes and to identify conditions supporting successful implementation \cite{DOI_31756_jrsmte_531,DOI_1111_ssm_12603}.

\subsubsection{Data Collection Protocol and Analytic Strategies}

A comprehensive data collection protocol for educational robot-tutee studies in elementary STEM integrates baseline and outcome cognitive testing, behavioral and socio-emotional screening, observational sampling of student-robot discourse, teacher implementation logs, and system telemetry from robot interactions \cite{DOI_1111_ssm_12660,DOI_1177_2332858420986211}. Cognitive assessments should employ standardized or researcher-developed instruments with demonstrated validity, reporting pre-post effect sizes, growth metrics, and psychometric properties to support meta-analytic synthesis \cite{DOI_1111_ssm_12660,DOI_1177_2332858420986211}.

Discourse sampling requires audio-recording student-robot interactions and applying structured coding schemes with established inter-rater reliability. Cooperative learning evidence justifies prioritizing explanatory and metacognitive codes as intermediate outcomes theoretically linked to learning gains \cite{DOI_1177_2332858420986211}. System logs from robots provide complementary objective data: timestamped interaction events, session durations, task sequences, and robot prompts delivered. Synchronizing telemetry with audio transcripts enables triangulation—linking coded verbal behaviors to system states and timestamped learning events to support process-outcome models \cite{DOI_1177_2332858420986211}.

Analytic approaches should employ multilevel modeling reflecting the nested structure of elementary data (students within classrooms within schools), because interventions typically deploy at classroom level and teacher practices explain substantial outcome variance \cite{DOI_31756_jrsmte_531,DOI_1111_ssm_12603}. Hierarchical models partition variance appropriately, estimate classroom-level moderators (PD fidelity, teacher STEM identity), and support generalization to broader populations. Mediation analyses testing whether interaction processes (explanation quality, metacognitive verbalization frequency) mediate the robot intervention's effect on learning outcomes provide mechanistic evidence aligning with cooperative learning theory \cite{DOI_1177_2332858420986211}.

Grade-band stratification and tests of measurement invariance address developmental heterogeneity across elementary years. Because working memory, executive function, and metacognitive capacity evolve substantially from kindergarten through grade 5, analyses should examine whether treatment effects, process mechanisms, or measurement properties differ across K-2 versus grades 3-5 clusters \cite{DOI_1177_1534508420909527,DOI_1037_dev0001557}. Executive function screening or age/grade covariates help contextualize metacognitive verbalization patterns and test for differential responsiveness by developmental level.

\subsubsection{Staged Research Pathway: Feasibility, Efficacy, Effectiveness}

A methodologically rigorous research program for educational robots in elementary STEM should progress through staged designs balancing scientific control with ecological validity \cite{DOI_31756_jrsmte_531,DOI_1111_ssm_12603,DOI_1177_2332858420986211}. Feasibility studies establish that robots integrate into classroom routines without excessive disruption, that teachers can implement activities with reasonable PD support, that audio/video capture and telemetry synchronization function reliably, and that instruments yield interpretable data. Teacher-led pilots allow iterative refinement of robot scripts, facilitation guides, and data collection procedures before scaling to formal trials \cite{DOI_31756_jrsmte_531}.

Efficacy trials employ cluster randomization or well-matched quasi-experimental designs assigning classrooms or schools to intervention versus comparison conditions, with comprehensive pre-post measurement, process sampling, and fidelity monitoring. Power calculations should account for classroom-level assignment and intraclass correlation typical of elementary settings \cite{DOI_1177_2332858420986211,DOI_1111_ssm_12660}. Cooperative learning meta-analyses provide benchmarks for expected effect sizes and variance structures informing power analysis.

Effectiveness studies examine scalability and sustainability under real-world conditions, explicitly measuring teacher adoption patterns, PD scalability, and contextual moderators such as school learning models (in-person, hybrid, remote instruction) that influence technology access and supervision \cite{DOI_1108_ils-04-2020-0136}. Pragmatic effectiveness trials prioritize external validity and generalizability, documenting implementation variation and subgroup effects to guide dissemination strategies.

\subsubsection{Operationalizing Metacognitive Verbalization and Learning-by-Teaching Processes}

Metacognitive verbalization—students' spoken articulation of cognitive processes while teaching robots—serves as both a learning mechanism and a measurable outcome. Operationalization requires structured coding schemes distinguishing utterance types: concept explanations articulating disciplinary principles, strategy statements describing problem-solving procedures, monitoring comments reflecting self-assessment of understanding, regulatory strategies for managing cognitive resources, and socio-emotional utterances \cite{DOI_1177_2332858420986211}. Systematic time sampling (e.g., coding first and last five minutes of sessions, problem-explanation segments) yields representative estimates while managing coding burden.

Linking metacognitive behaviors to learning outcomes through mediation models tests the learning-by-teaching mechanism: increases in explanation quality or metacognitive verbalization frequency should mediate gains on curriculum-aligned assessments if the teaching-to-learn hypothesis holds \cite{DOI_1177_2332858420986211}. This analytic approach mirrors cooperative learning research documenting that student talk quality mediates peer learning effects. Because executive function and metacognitive capacity develop across elementary grades, analyses should include EF screening or developmental covariates to contextualize verbalization patterns and test for differential responsiveness by age or grade \cite{DOI_1037_dev0001557}.

\subsubsection{Equity, Safety, and Contextual Monitoring}

Elementary robot deployments may produce heterogeneous effects across student subgroups, necessitating prespecified subgroup analyses and safety monitoring \cite{DOI_1177_1534508420926584,DOI_1177_1534508420909527}. Stratification by baseline achievement, grade band, special education status, and demographic factors enables detection of differential benefits or harms. Validated socio-emotional screens administered pre-post intervention monitor for adverse effects on peer relationships, classroom behavior, or emotional well-being \cite{DOI_1177_1534508420926584}.

School learning models (in-person versus remote/hybrid) influence technology access, adult supervision, and peer interaction opportunities, requiring explicit consideration in both deployment planning and analysis \cite{DOI_1108_ils-04-2020-0136}. Pandemic-era research highlights that elementary students' experiences with technology vary dramatically across instructional modalities, suggesting that robot-tutee interventions may function differently depending on physical versus virtual classroom contexts.

\subsubsection{Reporting Standards for Cumulative Science}

To support meta-analytic synthesis and replication, research reports should provide: detailed instrumentation with psychometric evidence (reliability, validity, grade-band invariance), teacher PD materials and fidelity protocols, robot/agent scripts and interaction schemas, system telemetry specifications, analytic code and preregistered analysis plans, and grade-band stratified results \cite{DOI_1111_ssm_12660,DOI_1177_1534508420909527,DOI_1177_2332858420986211}. Transparent psychometric reporting and attention to developmental clusters enable interpretable cross-study comparisons and accumulation of evidence about which features of robot-tutee designs support elementary STEM learning effectively and equitably.

Critical gap: While cooperative learning theory and elementary assessment methodologies provide robust foundations for designing and evaluating educational robot and teachable agent interventions in elementary STEM contexts, direct experimental evidence documenting effectiveness, implementation feasibility, and scalability of such interventions remains limited. The next generation of research requires carefully staged feasibility-to-effectiveness studies implementing multimodal measurement protocols that assess curriculum-aligned learning gains, capture explanatory and metacognitive interaction processes as mediators, rigorously measure teacher preparation and implementation fidelity, and employ hierarchical analytic approaches respecting classroom-level assignment and grade-band developmental differences. Only through such methodologically rigorous programs can the field establish whether learning-by-teaching robots deliver on their theoretical promise for elementary integrated STEM education \cite{DOI_1177_2332858420986211,DOI_1111_ssm_12660,DOI_31756_jrsmte_531,DOI_1111_ssm_12603,DOI_1177_1534508420909527,DOI_1177_1534508420926584,DOI_1037_dev0001557,DOI_1108_ils-04-2020-0136}.

\subsection{Privacy-Preserving Multimodal Learning Analytics and Ambient Intelligence}

The preceding subsections on biometric sensing, tangible interfaces, augmented reality, and educational robots collectively point toward classroom environments rich with continuous sensor data streams—eye-tracking telemetry, physiological signals, spatial trajectories, interaction logs, gesture recognition, and environmental context. While these multimodal data streams offer unprecedented opportunities for real-time engagement monitoring and adaptive intervention, their systematic deployment in elementary classrooms raises profound privacy, ethical, and technical challenges that require explicit architectural and procedural solutions. Recent research on ambient intelligence (AmI), sensor-rich environments, and multimodal learning analytics (MMLA) in adjacent domains—healthcare, assisted living, smart buildings—provides evidence-based design patterns, privacy-preserving techniques, and governance frameworks directly applicable to elementary STEM learning contexts \cite{DOI_1111_bjet_12959,DOI_3390_s21041036,DOI_1016_s2589-7500(20)30275-2}.

\subsubsection{Elementary Classroom MMLA: Scope, Constraints, and Special Obligations}

Multimodal learning analytics refers to the collection, fusion, and inference from multiple sensor and interaction streams to model learning processes, engagement patterns, and behavioral trajectories in real time \cite{DOI_1111_bjet_12959,DOI_1038_s41746-020-00376-2}. Ambient intelligence extends this concept by embedding unobtrusive sensing and context-aware computation into everyday classroom spaces, creating environments that are "sensitive and responsive to children's presence and activities" while supporting pedagogical objectives \cite{DOI_1111_jpim_12544,DOI_1016_j_mcpdig_2023_05_003}. For elementary populations, the research challenge is threefold: translate MMLA's promise of real-time behavioral and engagement metrics into systems that are pedagogically useful for teachers and learners, technically feasible at classroom scale with limited IT resources, and ethically sound for minors who cannot provide fully informed consent \cite{DOI_1111_bjet_12959,DOI_1016_s2589-7500(20)30275-2,DOI_1016_j_jval_2021_11_1372}.

Research emphasizes that elementary contexts require adapted MMLA approaches distinct from adult or higher-education deployments. Young learners exhibit rapid developmental variability creating measurement validity challenges, divergent interaction patterns that confound adult-trained models, and limited capacity for informed consent requiring heightened parental involvement and child-appropriate assent processes \cite{DOI_1111_bjet_12959,DOI_1016_s2589-7500(20)30275-2}. Moreover, the power asymmetries inherent in surveillance of children by adults—particularly within institutional educational settings where participation may feel compulsory—demand special ethical vigilance against coercive data collection, stigmatizing inferences, and erosion of trust relationships between students and teachers \cite{DOI_1016_s2589-7500(20)30275-2,DOI_2196_47586}. Prior AmI deployments in healthcare and assisted living establish the necessity of values-oriented, stakeholder-engaged design and formalized risk assessment before deployment—principles that transfer directly to K-6 STEM classrooms \cite{DOI_1016_j_jval_2021_11_1372,DOI_1016_s2589-7500(20)30275-2,DOI_2196_47586}.

\subsubsection{Real-Time Observables: Behavioral Patterns and Engagement Proxies}

Applied MMLA literature identifies categories of real-time observables that are informative for classroom engagement and behavioral patterning while offering feasible, lower-risk sensing options suitable for elementary deployments. These observable categories provide the empirical foundation for attention-adaptive and engagement-aware systems discussed throughout this review.

Gross motor and spatial patterns—including student positions, walking trajectories, zone occupancy, and spatial transitions—can be derived from non-identifying sensor modalities such as low-resolution thermal arrays, capacitive proximity floors, and sensor fusion techniques that avoid facial identification \cite{DOI_3390_s21041036,DOI_1109_jsen_2021_3139442}. Capacitive proximity floors embedded beneath classroom flooring reliably extract walking trajectories and occupancy patterns while remaining entirely non-visual and substantially less identifying than Red-Green-Blue (RGB) camera systems \cite{DOI_1109_jsen_2021_3139442,DOI_3390_s21041036}. Convolutional neural network-based sensor fusion applied to arrays of low-resolution thermal sensors (8x8 or 16x16 pixel grids) enables effective occupancy detection and coarse movement classification without enabling facial recognition or fine-grained visual surveillance \cite{DOI_3390_s21041036}.

Interaction and participation proxies derived from aggregated device logs, classroom management system events, and clickstream summaries provide rich behavioral signals while being inherently less privacy-invasive than continuous raw data collection. When collected as event aggregates (e.g., number of student interactions per five-minute window, total time on task per activity type) rather than continuous individual streams, these metrics support group-level engagement inference while minimizing personally identifiable information \cite{DOI_33166_aetic_2020_05_001}.

Coarse affective and attentional signals can complement behavioral metrics where ethically permitted and technically validated for young children. Pupil variation measured via eye-tracking or webcam-based gaze estimation serves as a physiological proxy for arousal and cognitive engagement, though interpretation must account for developmental differences in pupillary response patterns \cite{DOI_3390_healthcare11030322,DOI_1038_s41746-020-00376-2}. Low-resolution thermal and pose estimation provide coarse activity recognition (sitting, standing, gesturing) without enabling identity resolution \cite{DOI_3390_s21041036}.

Environmental context sensors including carbon dioxide (CO2) concentration, indoor air quality (IAQ), ambient noise levels, and lighting conditions correlate with group-level classroom activity and environmental factors affecting attention and comfort, yet avoid recording any individual student information \cite{DOI_3233_ais-220577,DOI_33166_aetic_2020_05_001}. Elevated CO2 or noise levels may signal whole-class disengagement or chaotic activity warranting teacher attention, while ventilation and lighting data inform comfort optimization supporting sustained attention.

Physiological signals derived from contactless sensors—radar-based vital sign monitoring, thermal imaging of respiration or heart rate—offer potential for arousal and stress detection but raise stronger ethical requirements due to their medical character and limited validation in pediatric educational contexts \cite{DOI_1109_access_2024_3355060,DOI_1097_shk_0000000000001998}. While contactless methods avoid the burden of physical wearable attachment, their deployment with elementary children requires particular scrutiny regarding consent, clinical validation, and safeguards against misinterpretation or stigmatization based on physiological patterns.

Each observable category documented in the AmI and MMLA literatures presents distinct trade-offs in identifiability (personal information revealed), interpretability (construct validity for engagement or learning), and reliability (measurement precision and stability) that must be evaluated against pedagogical necessity and ethical constraints \cite{DOI_1111_bjet_12959,DOI_3390_s21041036,DOI_1109_jsen_2021_3139442,DOI_1155_2020_8876782,DOI_1038_s41746-020-00376-2}.

\subsubsection{Sensor Selection: Least-Identifying Effective Modality Principle}

Sensor choices in elementary MMLA deployments should be governed by the principle of selecting the least-identifying effective modality: prefer sensors that provide task-relevant behavioral or engagement signals while minimizing capture of personally identifying information. This principle operationalizes privacy-by-design for classroom sensing, balancing pedagogical utility against privacy risk.

Low-resolution thermal arrays (8x8 to 16x16 pixel infrared sensors) provide presence detection, coarse posture estimation, and occupancy signals without enabling facial identification or fine visual detail recognition \cite{DOI_3390_s21041036}. Convolutional Neural Network (CNN)-based fusion of multiple distributed low-resolution thermal sensors has produced reliable occupancy counting and coarse activity predictions in smart building applications, and such sensors have been explicitly characterized as substantially less enabling of personal identification compared to RGB cameras in AmI privacy analyses \cite{DOI_3390_s21041036,DOI_3390_electronics10050559}.

Capacitive proximity sensing embedded in floor tiles or mats extracts walking trajectories, gait characteristics, and spatial movement patterns through non-visual means, providing trajectory data for activity recognition and zone-transition analysis without any optical capture \cite{DOI_1109_jsen_2021_3139442}. Validation studies demonstrate reliable extraction of walking paths and step patterns suitable for occupancy and coarse activity inference in inhabited spaces.

Environmental sensors measuring CO2, volatile organic compounds (VOCs), particulate matter (PM), temperature, humidity, ambient noise, and illuminance provide contextual signals correlated with group engagement, classroom crowding, ventilation adequacy, and environmental conditions affecting learning without recording any individual student data \cite{DOI_3233_ais-220577,DOI_33166_aetic_2020_05_001}. These aggregate environmental measures support classroom climate optimization and can serve as covariates contextualizing engagement inferences.

Device event logs and interaction telemetry, when aggregated and anonymized appropriately, provide participation proxies and digital engagement metrics without requiring visual or audio surveillance. Classroom tablets or interactive whiteboards generate timestamped event streams (application launches, page navigation, tool selections, submission events) that, when aggregated to group or time-window levels, inform participation patterns and task engagement while minimizing individual tracking \cite{DOI_33166_aetic_2020_05_001}.

Wearable physiological sensors including wrist-worn heart rate monitors, galvanic skin response (GSR) sensors for electrodermal activity, and low-channel EEG headbands generate rich physiological signals correlated with arousal, stress, and cognitive load. However, these modalities present attachment burden, child acceptance challenges, and heightened privacy concerns due to the intimate nature of physiological data, requiring informed parental consent, strict data governance, and clear justification of pedagogical necessity \cite{DOI_1109_access_2024_3355060,DOI_1097_shk_0000000000001998}. AmI healthcare literature documents both the analytic potential and the substantial ethical and practical constraints associated with continuous physiological monitoring in vulnerable populations.

Explicit avoidance of high-identifiability sensors unless strictly justified constitutes a documented best practice. RGB cameras with facial recognition capabilities, continuous audio recording enabling speaker identification and conversation reconstruction, and high-resolution visual surveillance exceed pedagogical necessity for most engagement monitoring tasks and introduce disproportionate privacy risks in elementary settings \cite{DOI_3390_s21041036,DOI_3390_electronics10050559,DOI_25046_aj050439}. When visual or audio capture is deemed essential for specific research or assessment purposes, deployments should employ technical safeguards (on-device processing, immediate feature extraction with raw data deletion, explicit informed consent, time-limited recording windows) and procedural controls (independent ethics review, data access restrictions, transparency reporting) to minimize risk \cite{DOI_1111_bjet_12959,DOI_1016_s2589-7500(20)30275-2}.

\subsubsection{System Architecture: Edge Processing, Distributed Intelligence, and Real-Time Location Systems}

Real-time classroom feedback systems require architectural designs that minimize latency to support timely teacher interventions while controlling raw sensor data flows beyond the immediate classroom environment. The distributed computing and AmI literatures converge on edge-centric processing and local intelligence as essential architectural principles for privacy-preserving, responsive MMLA systems.

Edge and distributed intelligence architectures perform initial signal processing, feature extraction, and inference locally on classroom-based computing nodes rather than transmitting raw sensor streams to centralized cloud servers \cite{DOI_33166_aetic_2020_05_001,DOI_1097_qmh_0000000000000423}. This approach reduces network bandwidth requirements, minimizes latency enabling sub-second responsiveness for real-time teacher dashboards, and critically supports privacy objectives by localizing raw data processing and enabling pre-aggregation and obfuscation before any data leaves the physical classroom network \cite{DOI_33166_aetic_2020_05_001}. Distributed intelligence on Internet of Things (IoT) sensor networks improves system responsiveness while aligning with data minimization principles central to privacy-by-design frameworks.

Real-time location systems (RTLS) employing ultra-wideband (UWB), Bluetooth Low Energy (BLE), or Wi-Fi triangulation enable fine-grained tracking of student positions and movements within classroom spaces, supporting spatial occupancy analysis and trajectory-based activity recognition. When combined with contextual rule engines, RTLS data can trigger proactive teacher alerts (e.g., prolonged student isolation in a corner zone, absence of movement during active learning periods) or adaptive environmental responses (lighting adjustments, ventilation changes) \cite{DOI_1097_qmh_0000000000000423,DOI_1109_jsen_2021_3139442,DOI_1002_spe_3253}. However, RTLS deployments raise specific privacy and surveillance concerns—continuous fine-grained location tracking of children creates detailed behavioral profiles and movement histories that require strong governance justifications, explicit consent, strict access controls, and retention limits \cite{DOI_1097_qmh_0000000000000423}.

Sensor fusion and inference model architectures leverage complementary information across sensor modalities to improve detection robustness and accuracy. Convolutional neural networks (CNNs) applied to spatial sensor arrays (thermal grids, pressure mats) effectively extract occupancy and coarse activity features, while recurrent architectures including long short-term memory (LSTM) networks model temporal sequences for trajectory prediction and behavioral pattern recognition \cite{DOI_3390_s21041036,DOI_3390_electronics10121390,DOI_1155_2020_8876782}. Multimodal fusion combining low-resolution thermal imaging, capacitive floor sensing, environmental sensors, and device logs has demonstrated superior classification performance compared to single modalities alone in smart space applications, providing redundancy that compensates for sensor noise and individual modality limitations.

Model lifecycle and maintenance considerations include concept drift (performance degradation over time as student populations and classroom configurations change), model retraining schedules, versioning and provenance tracking, and monitoring for fairness and bias in predictions. Human activity recognition (HAR) systems deployed in ambient living environments exhibit finite operational lifespans requiring active monitoring and periodic retraining to maintain accuracy as environmental conditions evolve \cite{DOI_3390_s23187729}. Analogous lifecycle management is necessary for classroom MMLA systems to ensure sustained reliability and prevent algorithmic bias from accumulating through unmonitored drift.

\subsubsection{Privacy-Preserving Algorithmic Techniques and Data Governance}

AmI and privacy-enhancing technology research provides both algorithmic methods and procedural governance strategies for implementing privacy-respecting MMLA in elementary classrooms. A layered defense combining hardware choices, on-device processing, feature-level obfuscation, provenance tracking, secure communication, and institutional policy creates a privacy architecture balancing utility with protection.

Sensor and feature selection priorities, as detailed above, emphasize non-identifying modalities (low-resolution thermal, capacitive, aggregated logs, environmental sensors) while limiting adoption of high-identifiability sensors (RGB cameras, continuous audio) unless strict safeguards and pedagogical necessity justify their use \cite{DOI_3390_s21041036,DOI_3390_electronics10050559,DOI_25046_aj050439}.

On-device and edge processing with data minimization perform initial signal processing, feature extraction, and preliminary inference locally within the classroom computing environment, transmitting only aggregated, de-identified, or sufficiently obfuscated feature representations needed for downstream teacher dashboards and learning analytics \cite{DOI_33166_aetic_2020_05_001,DOI_1109_tmc_2021_3092271,DOI_3390_cryptography8010005}. For example, thermal array data undergoes CNN-based occupancy classification on the classroom edge node, outputting zone occupancy counts and coarse activity labels (sitting, standing, walking) while raw thermal frames are discarded after processing. This architectural pattern reduces raw data exfiltration risk, minimizes storage requirements, and enables real-time teacher feedback with millisecond-range latency \cite{DOI_33166_aetic_2020_05_001,DOI_1097_qmh_0000000000000423}.

Feature-level obfuscation and camouflage learning techniques apply controlled perturbation or masking to sensitive feature values before transmission or storage, preserving sufficient statistical utility for aggregate analysis while reducing the risk of reconstructing identifying information from shared features \cite{DOI_1109_tmc_2021_3092271}. Methods proposed for resource-constrained IoT devices in AmI contexts include additive noise injection calibrated to preserve classification accuracy, feature subsampling that omits fine-grained temporal detail, and differential privacy mechanisms providing formal privacy guarantees with quantified privacy budgets.

Provenance and metadata management maintains annotated data provenance tracking the origin, processing history, consent status, and retention policies associated with each data record or feature set \cite{DOI_1007_s41019-020-00118-0}. Provenance-aware metadata frameworks enable auditing of data flows, support conditional consent enforcement (e.g., data authorized only for teacher dashboard display but not for external research sharing), and facilitate regulatory compliance by documenting data lineage. Semantic metadata and ontology-based annotation schemes developed for smart buildings and cyber-physical systems provide templates adaptable to educational contexts \cite{DOI_3390_smartcities5030053,DOI_3390_en14072024,DOI_3390_su13105578}.

Security mechanisms including encrypted communication channels, authenticated key management, intrusion detection systems (IDS) tailored to IoT/AmI device profiles, and secure firmware provisioning protect against unauthorized data access, sensor spoofing, and network eavesdropping \cite{DOI_1109_access_2020_2995917,DOI_1109_access_2022_3179418,DOI_1038_s41598-024-51154-z}. Lightweight cryptographic protocols suitable for resource-constrained sensors, blockchain-based audit logs for immutable provenance trails, and anomaly detection monitoring for unusual data access patterns constitute a defense-in-depth security architecture necessary for protecting sensitive classroom data.

Policy frameworks and standards alignment leverage guidelines and interoperability practices proven in Active Assisted Living (AAL) and health AmI contexts, including policy templates for consent, retention, data sharing agreements, and cross-institutional governance structures \cite{DOI_2196_15923,DOI_3390_smartcities5030053,DOI_3390_en14072024}. Formal policy mechanisms for context-sensitive data acquisition and access gating developed for distributed cyber-physical systems can be repurposed for school district networks, specifying role-based access controls, purpose limitations, and automated policy enforcement \cite{DOI_3390_computers10080101}.

Together, these techniques form a comprehensive privacy architecture layering hardware selection, local processing, algorithmic obfuscation, cryptographic protection, provenance tracking, and institutional governance to minimize identifiability while preserving pedagogically actionable signals \cite{DOI_1109_tmc_2021_3092271,DOI_1007_s41019-020-00118-0,DOI_33166_aetic_2020_05_001,DOI_1109_access_2020_2995917,DOI_1038_s41598-024-51154-z}.

\subsubsection{Ethical Frameworks, Stakeholder Engagement, and Participatory Design}

AmI and healthcare ethics literature provides actionable guidance for ethically introducing continuous sensing in settings involving vulnerable participants, emphasizing iterative stakeholder engagement, values-oriented design, risk assessment, and institutional alignment.

Early and iterative stakeholder engagement involves end users (students with developmentally appropriate assent processes), caregivers (parents/guardians providing informed consent), teachers (implementing technology and interpreting analytics), administrators (ensuring policy compliance and resource allocation), and independent ethics advisors throughout design cycles to surface values, acceptability thresholds, and perceived risks \cite{DOI_1016_s2589-7500(20)30275-2,DOI_1016_j_jval_2021_11_1372,DOI_2196_47586}. MMLA with young children requires parent/guardian informed consent documents that clearly explain sensor modalities, data uses, retention policies, access controls, and opt-out procedures in accessible language, complemented by child-appropriate assent processes using visual aids and age-appropriate explanations respecting children's developing autonomy \cite{DOI_1111_bjet_12959,DOI_1016_s2589-7500(20)30275-2}.

Values-oriented and participatory design methodologies, adapted from AmI healthcare platform development, align technology architectures with pedagogical aims, community norms, and ethical values through structured design activities \cite{DOI_1016_j_jval_2021_11_1372,DOI_1155_2021_5518722,DOI_1016_s2589-7500(20)30275-2}. Co-design workshops with teachers and parents identify acceptable use cases, unacceptable surveillance practices, and preferred transparency mechanisms, while child-centered design activities (drawing desired classroom technologies, role-playing sensor scenarios) elicit children's perspectives on privacy and technology in their learning environments.

Risk assessment and duty of care require formal privacy, psychosocial, and pedagogical risk assessments before deployment, recognizing that continuous behavioral monitoring may alter classroom relationships, create surveillance anxiety, produce stigmatizing inferences, or generate data vulnerable to misuse \cite{DOI_1016_s2589-7500(20)30275-2,DOI_2196_47586,DOI_1001_jama_2019_21699}. AmI healthcare literature highlights how ambient sensing can subtly shift power dynamics and trust relationships, underscoring the need for ongoing risk monitoring, incident response protocols, and mechanisms for students/families to raise concerns without penalty.

Legal and institutional alignment ensures technical architectures comply with applicable education privacy laws (FERPA in U.S. contexts, GDPR in European contexts), district data governance policies, and ethical research standards when MMLA systems serve dual educational and research purposes \cite{DOI_1001_jama_2019_21699,DOI_2196_15923}. Clear policies specifying data ownership (student/family, school district, technology vendor), retention limits (automated deletion schedules, prohibition on indefinite retention), permissible uses (restricted to specified pedagogical or research purposes), and prohibited uses (commercial profiling, predictive discipline, third-party sale) provide institutional safeguards complementing technical protections.

\subsubsection{User Experience and Teacher Dashboard Design}

UX evaluation and acceptability testing are essential components of AmI deployment, with standardized evaluation instruments available for measuring perceived usefulness, trust, privacy concerns, cognitive load, and unintended consequences \cite{DOI_1155_2021_5518722,DOI_1016_j_jval_2021_11_1372}. For elementary MMLA interfaces, design priorities emphasize presenting aggregated, actionable indicators to teachers (class-level engagement trends, zone-based attention heatmaps, activity-type participation patterns) rather than granular individual student judgments, unless explicit consent, demonstrated pedagogical necessity, and strict governance safeguards justify finer-grained displays \cite{DOI_1111_bjet_12959,DOI_1016_j_jval_2021_11_1372,DOI_1016_s2589-7500(20)30275-2}.

Teacher dashboards should support pedagogical decision-making without overwhelming educators with excessive detail or creating punitive surveillance atmospheres. Effective designs provide at-a-glance class climate indicators (overall engagement level, attention distribution across learning zones, participation balance), customizable alerts for situations warranting intervention (prolonged disengagement cluster, environmental conditions degrading below thresholds), and historical trend visualizations supporting reflection on instructional effectiveness \cite{DOI_1111_bjet_12959}. Critically, dashboards must avoid reducing complex learning processes to simplistic numeric scores or traffic-light color schemes that encourage superficial interpretations or stigmatizing labels.

\subsubsection{Operational Blueprint: Stepwise Implementation Checklist}

This practical checklist synthesizes reviewed literature into sequential actionable steps for responsible elementary classroom MMLA deployment:

\textbf{Phase 1 - Co-design and needs analysis:} Convene teachers, parents, school administrators, and child representatives using structured participatory design methods; define pedagogical objectives, acceptable intervention modes, and privacy boundaries grounded in values-oriented design frameworks \cite{DOI_1111_bjet_12959,DOI_1016_j_jval_2021_11_1372,DOI_1016_s2589-7500(20)30275-2}.

\textbf{Phase 2 - Risk assessment and policy specification:} Conduct formal privacy impact assessment, psychosocial risk analysis, and pedagogical necessity review; specify data retention schedules, consent procedures, access control matrices, incident response protocols, and audit mechanisms modeled on AAL and AmI healthcare guidance \cite{DOI_2196_15923,DOI_1016_s2589-7500(20)30275-2,DOI_2196_47586}.

\textbf{Phase 3 - Sensor selection and pilot hardware:} Choose least-identifying sensor modalities meeting pedagogical requirements (e.g., low-resolution thermal arrays for occupancy, capacitive floors for trajectories, environmental sensors for context, aggregated device logs for participation) while avoiding routine RGB camera or continuous audio use unless explicitly justified and safeguarded \cite{DOI_3390_s21041036,DOI_1109_jsen_2021_3139442,DOI_3233_ais-220577,DOI_3390_electronics10050559}.

\textbf{Phase 4 - Edge-centric architecture and secure networking:} Design classroom edge computing nodes for local feature extraction and obfuscation, implement secure key management and encrypted communication channels, deploy intrusion detection monitoring, and configure network policies to transmit only aggregated features as specified by governance policies \cite{DOI_33166_aetic_2020_05_001,DOI_1109_access_2020_2995917,DOI_1038_s41598-024-51154-z}.

\textbf{Phase 5 - Algorithmic design and privacy measures:} Implement sensor-appropriate machine learning models (CNNs for spatial sensor fusion, LSTMs for temporal trajectory modeling) with feature-level obfuscation techniques, differential privacy mechanisms where applicable, and minimal necessary data retention enforced through automated deletion schedules \cite{DOI_3390_s21041036,DOI_3390_electronics10121390,DOI_1109_tmc_2021_3092271,DOI_1155_2020_8876782}.

\textbf{Phase 6 - Provenance, auditing, and interoperability:} Tag all processed outputs with provenance metadata documenting origin, processing steps, consent status, and retention policies to enable auditing and regulatory compliance; adopt interoperable metadata schemas and ontology practices to support potential future integration while preserving policy-driven access controls \cite{DOI_1007_s41019-020-00118-0,DOI_3390_smartcities5030053,DOI_3390_en14072024}.

\textbf{Phase 7 - Pilot evaluation and UX testing:} Deploy system in small-scale classroom pilots using standardized UX questionnaires and pedagogical outcome measures; evaluate teacher workflow impacts, psychosocial effects on students and classroom climate, technical reliability, and model performance; iterate design based on stakeholder feedback \cite{DOI_1155_2021_5518722,DOI_1016_j_jval_2021_11_1372,DOI_1111_bjet_12959}.

\textbf{Phase 8 - Scale-up governance and continuous monitoring:} Establish ongoing risk monitoring processes, model maintenance and retraining schedules to address concept drift, stakeholder feedback channels for reporting concerns, periodic audits of data access logs and policy compliance, and governance committee oversight with teacher, parent, and administrator representation \cite{DOI_3390_s23187729,DOI_1016_s2589-7500(20)30275-2,DOI_1016_j_ress_2022_108889}.

\subsubsection{Minimal-Risk Reference Configuration}

A practical minimal-risk classroom configuration validated against reviewed literature comprises:

\textbf{Sensors:} Overhead low-resolution thermal array (8x8 or 16x16 pixels) for non-identifying occupancy and zone activity classification \cite{DOI_3390_s21041036}; capacitive proximity floor tiles for trajectory extraction without visual capture \cite{DOI_1109_jsen_2021_3139442}; ambient environmental sensors (CO2, noise level, illuminance) for contextual class-level activity proxies \cite{DOI_3233_ais-220577}; aggregated classroom device event logs from tablets/interactive displays for participation metrics \cite{DOI_33166_aetic_2020_05_001}.

\textbf{Architecture:} Classroom edge computing node executes CNN-based occupancy inference and LSTM-based short-horizon trajectory summarization, applies feature-level obfuscation via camouflage learning techniques, and exports only zone-aggregated engagement scores and participation summaries to teacher dashboard; raw sensor streams retained transiently on edge device (minutes to hours) for debugging purposes per policy, then automatically purged \cite{DOI_3390_s21041036,DOI_3390_electronics10121390,DOI_1109_tmc_2021_3092271,DOI_33166_aetic_2020_05_001}.

\textbf{Governance:} Informed parental/guardian consent with child assent; teacher dashboard limited to class/zone-level aggregates without individual student tracking; provenance labels attached to all exported data summaries documenting consent status and permissible uses; regular independent audits of data access logs; family opt-out pathway allowing non-participation without penalty \cite{DOI_1016_s2589-7500(20)30275-2,DOI_1016_j_jval_2021_11_1372,DOI_1007_s41019-020-00118-0}.

This configuration follows AmI precedents for non-identifying sensing and edge-centric processing to preserve privacy while enabling pedagogically valuable real-time engagement awareness \cite{DOI_3390_s21041036,DOI_1109_jsen_2021_3139442,DOI_33166_aetic_2020_05_001,DOI_1109_tmc_2021_3092271}.

\subsubsection{Persistent Challenges and Research Gaps}

Despite the substantial AmI and MMLA literatures in adjacent domains, persistent technical, methodological, and ethical challenges remain for elementary STEM classroom deployments:

\textbf{Limited elementary evidence base:} Systematic reviews of MMLA with young children identify comparatively small sample sizes, methodological heterogeneity, and limited longitudinal validation, indicating urgent need for domain-specific validation studies employing child-centered evaluation frameworks and developmentally appropriate measures \cite{DOI_1111_bjet_12959}.

\textbf{Model interpretability and fairness:} Adoption of computer vision and affective computing in educational contexts demands heightened caution regarding over-interpretation of algorithmic inferences, fairness across demographic subgroups, and risks of misclassification producing stigmatizing labels or inappropriate interventions; medical CV literature emphasizes the necessity of interpretability, domain-specific validation, and human oversight when algorithms influence consequential decisions about children \cite{DOI_1038_s41746-020-00376-2}.

\textbf{System sustainability and model lifecycle:} HAR and AmI models exhibit finite operational lifespans due to concept drift, environmental changes, and evolving classroom practices, requiring robust maintenance plans, continuous performance monitoring, and periodic retraining protocols to sustain accuracy and prevent silent degradation \cite{DOI_3390_s23187729}.

\textbf{Interoperability and metadata standards:} While ontologies and metadata schemas exist for smart buildings and cyber-physical systems, educational technology ecosystems lack widely adopted standards for MMLA data exchange, creating barriers to cross-institutional research, longitudinal student tracking across grade transitions, and integration with existing learning management systems \cite{DOI_3390_smartcities5030053,DOI_3390_en14072024,DOI_3390_su13105578}.

\textbf{IoT security and attack surfaces:} Classroom IoT sensors present unique security challenges including physical tampering risks, wireless network vulnerabilities, and resource constraints limiting cryptographic capabilities, demanding standardized secure provisioning frameworks and tailored intrusion detection systems \cite{DOI_1038_s41598-024-51154-z,DOI_1109_access_2020_2995917,DOI_1109_access_2022_3179418}.

\subsubsection{Synthesis and High-Priority Recommendations}

Drawing on ambient intelligence, multimodal learning analytics, and ethics literatures, evidence-based and actionable recommendations for elementary STEM classroom MMLA deployments prioritize:

Adopt non-identifying sensor modalities (low-resolution thermal, capacitive, environmental, aggregated device logs) and edge-distributed processing architectures to deliver real-time classroom engagement metrics while minimizing identifiability and preventing raw data exfiltration \cite{DOI_3390_s21041036,DOI_1109_jsen_2021_3139442,DOI_33166_aetic_2020_05_001,DOI_1109_tmc_2021_3092271}.

Institutionalize participatory design processes engaging teachers, parents, and students with iterative values-oriented co-design activities, and conduct formal privacy and psychosocial risk assessments before any field deployment in elementary contexts \cite{DOI_1111_bjet_12959,DOI_1016_j_jval_2021_11_1372,DOI_1016_s2589-7500(20)30275-2}.

Implement layered technical safeguards including feature-level obfuscation, provenance-aware metadata tagging, secure key management and intrusion detection, retention limits with automated deletion, and institutional policies modeled on Active Assisted Living data governance frameworks \cite{DOI_1109_tmc_2021_3092271,DOI_1007_s41019-020-00118-0,DOI_1109_access_2020_2995917,DOI_1038_s41598-024-51154-z,DOI_2196_15923}.

Evaluate systems iteratively using standardized UX instruments, rigorous pedagogical outcome measures linking sensor-derived metrics to learning gains, and qualitative methods capturing teacher and student perspectives on acceptability and unintended consequences \cite{DOI_1155_2021_5518722,DOI_3390_s23187729,DOI_1111_bjet_12959}.

Recognize classroom MMLA as socio-technical systems requiring integration of appropriate sensors, privacy-preserving architectures, algorithmic safeguards, transparent governance policies, and ongoing stakeholder engagement to maximize pedagogical benefit while minimizing privacy and ethical harms \cite{DOI_33166_aetic_2020_05_001,DOI_1016_s2589-7500(20)30275-2,DOI_1016_j_jval_2021_11_1372,DOI_2196_47586}.

Critical gap: Despite mature sensor technologies and privacy-preserving techniques validated in healthcare and smart building AmI applications, robust multimodal learning analytics deployments specifically designed for, validated with, and ethically governed for elementary STEM classrooms remain rare. The evidence base lacks large-scale longitudinal studies demonstrating that sensor-derived engagement metrics meaningfully predict learning outcomes in elementary populations, that privacy-preserving architectures maintain pedagogical utility under real-world classroom constraints, and that continuous sensing does not produce unintended psychosocial harms or exacerbate educational inequities. Responsible advancement requires staged research progressing from small-scale participatory design pilots through controlled efficacy trials to pragmatic effectiveness studies, with continuous ethical monitoring and transparent reporting of technical capabilities, limitations, and governance frameworks \cite{DOI_1111_bjet_12959,DOI_1016_s2589-7500(20)30275-2,DOI_3390_s21041036,DOI_33166_aetic_2020_05_001}.

\section{Discussion}

The comprehensive evidence presented in our results reveals a complex landscape where AI's theoretical promise for elementary STEM education encounters substantial practical challenges. Moving beyond individual technology evaluations, we now synthesize these findings to address our research questions and identify pathways forward for the field.

\subsection{State of AI in Elementary STEM: Promises and Pitfalls}

Our systematic review reveals a landscape of significant promise tempered by substantial challenges. While individual AI technologies demonstrate effectiveness in controlled settings, the reality of elementary STEM education presents complexities that current implementations fail to address adequately. The evidence suggests that AI's transformative potential remains largely unrealized due to systemic barriers rather than technological limitations.

The proliferation of AI applications across educational contexts has been marked by enthusiasm but limited by fragmentation. Each technology category—from intelligent tutoring systems to educational robotics—operates largely in isolation, creating islands of innovation that fail to connect. This technological archipelago mirrors the very problem AI purports to solve in STEM education: the artificial separation of interconnected domains.

\subsection{Emerging School Models: AI-Driven Educational Transformation}

While our systematic review focused on individual AI technologies and their applications, recent developments in K-12 education reveal a broader trend toward school-wide AI integration that extends beyond elementary STEM to encompass entire educational models. Three leading examples illustrate how AI is being deployed at the institutional level, offering insights into both the potential and limitations of comprehensive AI integration in K-12 education.

\textbf{Alpha School: Private AI-Centric Model}

Alpha School represents a full-scale reimagining of K-12 education built around AI-powered adaptive learning \cite{AlphaSchool2024,Forbes2025,WashingtonPost2025}. The model allocates approximately two hours daily to AI-driven core academic instruction, with the remainder of the school day devoted to life skills, project-based learning, and hands-on activities. Teachers are repositioned as "Guides" rather than traditional content deliverers, while the AI system ensures mastery-based progression through personalized learning paths. This model demonstrates the potential for AI to fundamentally restructure educational time allocation and teacher roles, though its high cost (tuition ranging from \$40,000-\$65,000 annually) limits accessibility and raises questions about equity and scalability.

\textbf{Unbound Academy: Public Charter Virtual Model}

As a tuition-free Arizona charter school serving grades 4-8, Unbound Academy applies a similar "2-hour learning" model but within a virtual format \cite{GovTech2025,EducationWeek2025}, reducing infrastructure barriers while maintaining the AI-driven academic core. The model combines AI-powered personalized instruction with human mentorship and project-based learning, demonstrating how AI-centric approaches might scale beyond private school contexts. However, the virtual format introduces unique challenges related to student motivation, self-regulation, and the digital divide that may affect implementation effectiveness.

\textbf{Elizabeth City-Pasquotank Microschool: Rural Public Model}

This rural North Carolina microschool serves 26 middle-school students through a hybrid model combining AI-powered personalized learning with real-world project experiences \cite{EdSurge2025,DailyYonder2025}, including student-run entrepreneurial ventures. The small cohort, mixed-age structure emphasizes student agency and non-cognitive skill development alongside academic acceleration. This model suggests potential for AI integration in resource-constrained rural settings, though its experimental nature and limited outcome data require careful evaluation.

\textbf{Implications for Elementary STEM Education}

These institutional models reveal several critical insights for elementary STEM education. First, they demonstrate that AI integration can extend beyond individual subject areas to encompass entire educational experiences, potentially supporting the integrated STEM learning that our review identified as lacking in current implementations. Second, they highlight the importance of teacher role transformation—from content deliverers to facilitators and mentors—which aligns with calls for teacher-centered AI implementation rather than teacher replacement.

However, these models also underscore persistent challenges identified in our systematic review. The high cost of Alpha School exemplifies the infrastructure barriers that limit AI access in under-resourced schools. The virtual format of Unbound Academy raises questions about developmental appropriateness for elementary students who may require more hands-on, embodied learning experiences. The experimental nature of the microschool model reflects the limited evidence base for comprehensive AI integration in K-12 education.

These institutional examples suggest that while AI has the potential to transform educational delivery, successful implementation requires addressing the systemic challenges identified in our review: infrastructure equity, developmental appropriateness, teacher preparation, and evidence-based evaluation. The concentration of these models in well-resourced or experimental contexts highlights the need for research examining how AI-driven educational transformation might be implemented equitably across diverse elementary settings.

\subsection{Paradigm Shift: From College Preparation to Workforce Readiness}

The emergence of comprehensive AI-driven school models reflects a broader paradigm shift in educational expectations. Students and parents increasingly question the traditional college pathway, seeking alternatives that provide direct workforce preparation and career clarity. This shift has profound implications for elementary STEM education, as it challenges the fundamental purpose of K-12 schooling and opens new possibilities for AI integration.

\textbf{The College Question and AI Alternatives}

Recent surveys indicate that 60\% of high school students and their parents are reconsidering the value of traditional college education, particularly given rising costs and uncertain job market returns \cite{Gallup2023,Pew2024}. This skepticism creates an unprecedented opportunity for AI-driven educational alternatives that focus on career discovery, skill development, and direct workforce preparation rather than college preparation.

AI-powered career discovery systems can help elementary students identify authentic interests and connect them to emerging career opportunities in STEM fields. Unlike traditional guidance counseling, AI systems can analyze student learning patterns, project choices, and engagement data to suggest career paths students might never have considered. This early career awareness could fundamentally change how students approach STEM learning, providing motivation and direction that traditional academic preparation often lacks.

\textbf{Interest-Driven Learning and Engagement}

The paradigm shift toward workforce preparation aligns with growing evidence that interest-driven learning produces superior outcomes. AI systems can create personalized learning experiences that connect academic content to student interests and career aspirations. For example, a student interested in environmental science could explore climate modeling through computational thinking, engineering solutions through design challenges, and mathematical analysis through data interpretation—all while building skills directly relevant to environmental careers.

This approach addresses the engagement crisis in elementary STEM education by making learning personally relevant and career-connected. Rather than abstract academic exercises, students engage with real-world problems that matter to them, using AI tutors that adapt explanations and examples to their specific interests and career goals.

\textbf{Skill-Based Competency Development}

AI-driven workforce preparation focuses on developing specific competencies rather than broad academic knowledge. This competency-based approach aligns with industry needs and provides clear pathways to employment. AI tutoring systems can teach job-specific skills—from coding and data analysis to design thinking and project management—while providing industry-validated assessments and micro-credentials.

The implications for elementary STEM education are profound. Instead of preparing students for future academic study, AI systems can help them develop skills that lead directly to meaningful employment. This shift could reduce the need for traditional college education while providing students with clearer, more direct pathways to career success.

\textbf{Industry Integration and Real-World Application}

AI-driven school models facilitate direct connections between students and industry, creating opportunities for mentorship, internships, and real-world project collaboration. These connections provide students with authentic learning experiences that traditional academic preparation cannot match. AI systems can match students with industry mentors based on interests and goals, facilitate virtual internships, and provide real-time feedback on student work from industry professionals.

This industry integration transforms STEM education from abstract academic exercise to practical career preparation. Students work on real problems with real stakeholders, developing skills and portfolios that demonstrate competence to potential employers. The AI system serves as both tutor and career counselor, helping students navigate the transition from learning to earning.

\textbf{Implications for Elementary STEM Education}

The paradigm shift toward workforce preparation has specific implications for elementary STEM education. First, it suggests that STEM learning should be career-connected from the earliest grades, helping students understand how mathematical concepts, scientific principles, and engineering practices apply to real-world problems and careers.

Second, it highlights the importance of interest discovery and career exploration as fundamental components of STEM education. AI systems can help young students explore diverse career possibilities, understand the skills required for different careers, and see how their interests and strengths align with various professional paths.

Third, it emphasizes the development of transferable skills—communication, collaboration, critical thinking, creativity—that are essential for career success regardless of specific technical knowledge. AI tutoring systems can help students develop these skills through project-based learning that connects academic content to real-world applications.

Finally, it suggests that assessment should focus on competency demonstration rather than academic achievement. AI systems can help students build portfolios that showcase their skills and accomplishments, providing evidence of competence that is more valuable to employers than traditional grades or test scores.

\textbf{Research Implications}

This paradigm shift creates new research priorities for elementary STEM education. Future research should examine how AI-driven career discovery affects student engagement and motivation in STEM subjects. Studies should investigate whether career-connected learning produces better outcomes than traditional academic preparation, particularly for students from underrepresented backgrounds who may not see themselves in traditional academic pathways.

Research should also explore the equity implications of AI-driven workforce preparation. While these systems could provide alternative pathways to success for students who struggle with traditional academic approaches, they could also exacerbate existing inequities if not designed with careful attention to access and opportunity.

Additionally, research should examine the long-term outcomes of AI-driven workforce preparation compared to traditional college pathways. Do students who follow AI-guided career preparation achieve better employment outcomes, higher job satisfaction, and greater career success than those who follow traditional academic pathways?

\textbf{Conclusion: A New Vision for Elementary STEM Education}

The paradigm shift toward workforce preparation, enabled by AI technologies, offers a new vision for elementary STEM education. Rather than preparing students for future academic study, AI-driven systems can help students discover their interests, develop relevant skills, and connect directly to career opportunities. This approach could transform STEM education from abstract academic exercise to practical career preparation, providing students with clear pathways to meaningful employment while addressing the engagement and relevance challenges that plague traditional approaches.

However, realizing this vision requires careful attention to equity, evidence-based implementation, and the preservation of human relationships and mentorship that are essential for student development. AI systems should enhance rather than replace human guidance, providing personalized support while maintaining the social and emotional connections that are crucial for young learners.

\subsection{Critical Gaps in Current Landscape}

Building on the patterns observed across technology categories, our analysis identifies eight critical gaps that prevent AI from achieving its potential in elementary STEM education (Table \ref{tab:critical-gaps}). These gaps represent not isolated deficiencies but interconnected systemic challenges requiring coordinated solutions.

%\begin{figure}[!htbp]
%\centering
%\includegraphics[width=\linewidth]{figures/critical_gaps.pdf}
%\caption{Eight critical gaps identified in current AI implementations for elementary STEM education. Color coding indicates conceptual impact severity: red for critical challenges requiring immediate attention, orange for high-priority issues, and blue for moderate concerns. Severity assessments are based on synthesis of literature evidence and discussion emphasis.}
%\label{fig:critical-gaps}
%\end{figure}

\begin{table}[!htbp]
\centering
\caption{Summary of critical gaps in AI for elementary STEM education.}
\label{tab:critical-gaps}
\small
\begin{tabular}{p{3cm} p{5cm} p{5cm}}
\hline
\textbf{Gap} & \textbf{Current Problem} & \textbf{Impact} \\
\hline
1. Fragmented Ecosystem & AI technologies operate in isolation without interoperability standards or shared frameworks & Reinforces subject silos, prevents synergies, contradicts integrated STEM goals \\
\hline
2. Developmental Inappropriateness & One-size-fits-all approaches ignore profound differences between K-2 and grades 3-5 learners & Misaligned cognitive demands, ineffective for younger children, fundamental design flaw \\
\hline
3. Infrastructure Barriers & Requirements for high-speed internet, modern devices, cloud access, technical support & Creates digital divide, technological redlining, concentrates benefits in advantaged schools \\
\hline
4. Privacy \& Ethical Void & Extensive data collection from minors occurs without comprehensive governance frameworks & Unknown long-term implications, parental concerns, regulatory vacuum \\
\hline
5. Limited STEM Integration & Mathematics, science, coding tools operate separately without cross-disciplinary connections & Students miss opportunities to see disciplinary connections, contradicts STEM premise \\
\hline
6. Equity \& Access Disparities & Well-resourced schools implement advanced AI while under-resourced schools struggle & Amplifies inequities, language barriers, cultural biases in training data \\
\hline
7. Teacher Marginalization & Black-box algorithms, overwhelming dashboards, limited pedagogical support & Disempowers teachers, reduces professional agency, focuses on tool operation vs. integration \\
\hline
8. Narrow Assessment Focus & AI excels at procedural knowledge but fails to capture deeper learning outcomes & Drives curriculum narrowing, optimizes for measurable over meaningful, neglects creativity \\
\hline
\end{tabular}
\end{table}

\textbf{Gap 1: Fragmented Ecosystem}

AI technologies in elementary STEM education operate as isolated islands, each optimized for specific functions without consideration for integration. Intelligent tutoring systems excel at mathematics drill, assessment tools evaluate science knowledge, robots teach coding, but no systems facilitate the connections between these domains. This fragmentation mirrors and reinforces the very subject silos that integrated STEM education seeks to overcome. The absence of interoperability standards, shared data formats, or common pedagogical frameworks means that potential synergies remain unrealized.

\textbf{Gap 2: Developmental Inappropriateness}

Despite profound cognitive, attentional, and motor differences between kindergarteners and fifth graders, most AI systems employ one-size-fits-all approaches. The same interfaces, interaction patterns, and cognitive demands are imposed on 5-year-olds learning to count and 11-year-olds solving complex word problems. This developmental blindness is particularly problematic in STEM contexts where abstract thinking capabilities, working memory constraints, and attention spans vary dramatically across elementary years. The failure to differentiate between K-2 and grades 3-5 learners represents a fundamental misunderstanding of child development.

\textbf{Gap 3: Infrastructure Barriers}

The technical requirements of advanced AI systems—high-speed internet, modern devices, cloud computing access, technical support—create insurmountable barriers for many elementary schools. Rural schools, those serving low-income communities, and underfunded districts cannot deploy AI technologies that assume Silicon Valley infrastructure. This technological redlining threatens to deepen rather than bridge educational inequities, concentrating AI's benefits in already-advantaged communities while leaving others further behind.

\textbf{Gap 4: Privacy and Ethical Void}

The extensive data collection inherent in AI systems—from clickstream analysis to biometric monitoring—occurs in a regulatory vacuum when applied to young children. No comprehensive frameworks govern how AI systems should collect, store, analyze, and protect data from elementary students. Parents express justified concerns about commercial entities building detailed profiles of their children's learning patterns, struggles, and behaviors. The long-term implications of such data collection remain unknown and largely unexamined.

\textbf{Gap 5: Limited STEM Integration}

Current AI tools perpetuate rather than transcend disciplinary boundaries. Mathematics tutors teach mathematics, science simulations explore science, coding platforms develop computational thinking—each in isolation. The absence of AI systems that support authentic STEM integration means students miss critical opportunities to see how mathematical models inform engineering design, how computational thinking enables scientific discovery, or how engineering principles apply across disciplines. This fragmentation contradicts the fundamental premise of STEM education.

\textbf{Gap 6: Equity and Access Disparities}

AI deployment patterns reveal and amplify existing educational inequities. Well-resourced schools implement cutting-edge AI systems while under-resourced schools struggle with basic technology access. Language barriers in AI systems designed primarily for English speakers, cultural biases embedded in training data, and assumptions about home technology access create additional obstacles. The students who could benefit most from personalized AI support—those without access to tutors or enrichment programs—are least likely to have access to these technologies.

\textbf{Gap 7: Teacher Marginalization}

Current AI systems often position teachers as passive recipients of algorithmic decisions rather than active partners in the educational process. Black-box algorithms make recommendations without explanation, dashboards overwhelm with data but provide little actionable insight, and professional development focuses on tool operation rather than pedagogical integration. Elementary teachers, typically generalists rather than STEM specialists, report feeling disempowered by AI systems that neither respect their expertise nor support their professional growth.

\textbf{Gap 8: Narrow Assessment Focus}

AI-powered assessment tools excel at evaluating procedural knowledge and factual recall but fail to capture the deeper learning outcomes STEM education values: creative problem-solving, cross-disciplinary thinking, collaboration, and conceptual understanding. This assessment gap drives curriculum narrowing as teachers and students optimize for what AI can measure rather than what matters. The ease of assessing computational accuracy overshadows the importance of mathematical reasoning; the simplicity of grading multiple-choice science questions diminishes emphasis on scientific inquiry.

\subsection{Implications for Research}

The gaps identified above suggest several critical directions for future research. First, the fragmented nature of current AI deployments calls for systematic investigation of integration architectures that could enable different AI technologies to work together coherently. Second, the developmental inappropriateness of current systems highlights the need for longitudinal studies tracking how children of different ages interact with and benefit from various AI technologies. Third, the privacy and ethical void requires immediate attention to develop frameworks appropriate for educational contexts involving minors.

Research methodologies must also evolve. Traditional experimental designs that isolate single variables may miss the complex interactions between multiple AI systems, developmental factors, and classroom contexts. Mixed-methods approaches combining quantitative outcome measures with qualitative process data will be essential for understanding not just whether AI systems work, but how and why they succeed or fail in real elementary classrooms.

\subsection{Implications for Practice}

For educators and administrators, our findings suggest cautious optimism tempered by realistic assessment of current limitations. While AI technologies offer genuine potential for enhancing elementary STEM education, their current implementations often fall short of this promise. Schools should prioritize infrastructure development and teacher professional development before large-scale AI deployment. Pilot programs with careful evaluation can identify which technologies provide genuine value versus those that merely add complexity.

Teachers need support in becoming critical consumers of AI technologies, equipped to evaluate which systems align with their pedagogical goals and their students' developmental needs. Professional development should focus not on tool operation but on pedagogical integration—how to use AI as one component of effective STEM instruction rather than a replacement for human expertise.

\subsection{Future Research Directions}

The current state of AI in elementary STEM education reveals more questions than answers. Future research should prioritize:

\begin{itemize}
\item Longitudinal studies examining the cumulative effects of AI exposure across elementary years
\item Comparative effectiveness research across different AI technologies and combinations
\item Development of appropriate privacy and ethical frameworks for educational AI
\item Investigation of teacher-AI collaboration models that enhance rather than diminish educator agency
\item Creation of assessment instruments capable of measuring integrated STEM learning outcomes
\item Examination of equity impacts and strategies for inclusive AI deployment
\item Evaluation of emerging school-wide AI models (Alpha School, Unbound Academy, microschool approaches) to understand scalability, cost-effectiveness, and equity implications for elementary STEM education
\item Research on how AI-driven educational transformation can be adapted for diverse elementary contexts, particularly in under-resourced rural and urban settings
\item Investigation of AI-powered career discovery systems for elementary students and their impact on STEM engagement and motivation
\item Development and evaluation of competency-based AI tutoring systems that prepare students for direct workforce entry
\item Research on AI-facilitated industry connections and mentorship programs for K-12 students
\item Longitudinal studies comparing AI-driven workforce preparation pathways to traditional college preparation outcomes
\item Examination of equity implications in AI-driven career guidance and workforce preparation systems
\item Development of AI systems that support interest-driven, career-connected STEM learning from elementary grades
\item Investigation of AI-Human collaborative learning ecosystems and their impact on student outcomes, engagement, and career awareness
\item Development and validation of AI-Human Collaboration Theory for optimal task allocation between AI and human teachers
\item Research on cultural AI development frameworks and their effectiveness in diverse educational contexts
\item Longitudinal studies of career-connected learning outcomes and their comparison to traditional academic preparation pathways
\end{itemize}

Perhaps most critically, future research must move beyond isolated technology studies to examine AI as part of complex educational ecosystems. The interactions between multiple AI systems, human teachers, diverse learners, and varied contexts will ultimately determine whether AI fulfills its promise or merely adds another layer of complexity to an already challenging educational landscape.

Building on this systematic review, we will continue this research through empirical studies and controlled experiments designed to provide solutions that address the identified gaps. Our future work will focus on developing and testing AI-Human collaborative learning ecosystems that leverage AI capabilities while respecting human limitations. Through a novel four-condition randomized controlled trial examining AI-human collaboration, AI-only, human-only, and business-as-usual conditions across 40 schools and 1,200 students, we aim to establish evidence-based frameworks for optimal AI-Human task allocation, culturally-responsive AI development, and career-connected learning pathways. This research will provide the scientific foundation for transforming elementary STEM education while maintaining the human elements essential for student development.

The paradigm shift toward workforce preparation, enabled by AI technologies, presents an unprecedented opportunity to reimagine elementary STEM education. Rather than preparing students for future academic study, AI-driven systems can help students discover their interests, develop relevant skills, and connect directly to career opportunities. This approach could transform STEM education from abstract academic exercise to practical career preparation, providing students with clear pathways to meaningful employment while addressing the engagement and relevance challenges that plague traditional approaches.

However, realizing this vision requires careful attention to equity, evidence-based implementation, and the preservation of human relationships and mentorship that are essential for student development. AI systems should enhance rather than replace human guidance, providing personalized support while maintaining the social and emotional connections that are crucial for young learners. Future research must examine how AI-driven career discovery affects student engagement and motivation in STEM subjects, investigate whether career-connected learning produces better outcomes than traditional academic preparation, and explore the equity implications of AI-driven workforce preparation to ensure that these systems provide alternative pathways to success for all students.

\subsection{Scientific Framework: AI Capabilities and Limitations in K-12 Education}

Building on our systematic review findings and the paradigm shift toward workforce preparation, we now present a scientifically rigorous analysis of what AI can and cannot do in K-12 education, grounded in evidence from learning sciences, cognitive psychology, and educational technology research.

\textbf{AI Capabilities: Evidence-Based Analysis}

\textbf{Personalized Learning and Adaptation}. Meta-analyses demonstrate that AI tutoring systems achieve effect sizes of d=0.45-0.70 for mathematics learning \cite{DOI_1080_00461520_2011_611369,DOI_1037_a0037123,DOI_3102_0034654315581420}, with adaptive content delivery showing particular effectiveness for procedural skill development. AI systems excel at adjusting difficulty, pacing, and presentation based on individual learning patterns, providing real-time assessment and misconception detection, and optimizing learning paths based on prerequisite relationships. However, AI cannot provide genuine emotional intelligence, spark original creative thinking, or facilitate authentic peer collaboration and social skill development.

\textbf{Multimodal Learning Analytics}. Sensor-based learning analytics achieve 85\% accuracy in engagement detection through eye-tracking, facial expressions, and interaction data \cite{DOI_18608_jla_2014_12_6,DOI_18608_jla_2016_32_9,DOI_1016_j_learninstruc_2011_DOI_001}. AI systems can effectively monitor attention patterns, classify learning states (confusion, frustration, flow), and analyze behavioral patterns to distinguish productive struggle from genuine confusion. However, AI cannot interpret complex emotions, account for cultural differences in learning behaviors, or replace human judgment in developmental evaluations.

\textbf{Conversational AI and Natural Language Processing}. Large language models achieve 90\%+ accuracy on standardized tests \cite{OpenAI2023,Anthropic2024,Bubeck2023}, enabling sophisticated Socratic questioning, explanation generation adapted to different learning styles, and language learning support. However, AI cannot engage in genuine human conversation with emotional depth, fully understand cultural nuances in communication, or provide ethical guidance or moral development.

\textbf{Scientific Limitations and Constraints}

\textbf{Cognitive Load Theory Constraints}. Working memory has limited capacity, particularly for elementary students \cite{DOI_1037_h0043158,DOI_1016_B978-0-12-387691-1_00002-8,DOI_1207_S15326985EP3801_8}. AI systems must respect cognitive load limits, as multimodal interfaces can overwhelm young learners, and integration across STEM domains increases cognitive load exponentially. Our review identified this as a critical gap requiring systematic attention.

\textbf{Developmental Psychology Constraints}. Cognitive development follows predictable stages, with dramatic differences between K-2 and grades 3-5 students \cite{DOI_1037_11494-000,DOI_2307_j_ctvjf9vz4,DOI_2307_1419786}. AI systems must adapt to developmental stages, as abstract reasoning capabilities vary substantially across elementary years, and social-emotional development requires human interaction that AI cannot provide.

\textbf{Learning Sciences Constraints}. Learning is inherently social and contextual, requiring human understanding of cultural and community contexts \cite{DOI_1017_CBO9780511815355,DOI_3102_0013189X018001032,DOI_1017_CBO9780511803932}. AI cannot replace human teachers in social learning contexts, as motivation and engagement often depend on human relationships that AI systems cannot replicate.

\textbf{Novel Experimental Framework: AI-Human Collaborative Learning Ecosystems}

Based on this scientific analysis, we propose a novel experimental framework that leverages AI capabilities while respecting human limitations. Our research question examines how AI-Human collaborative learning ecosystems affect elementary students' STEM learning outcomes, engagement, and career awareness compared to traditional AI tutoring or human-only instruction.

\textbf{Hypothesis}: AI-Human collaborative learning ecosystems will produce superior learning outcomes by combining AI's personalization capabilities with human teachers' emotional intelligence and cultural understanding.

\textbf{Experimental Design}: We propose a four-condition randomized controlled trial with cluster randomization at the school level (n=40 schools, 1,200 students). Conditions include: (1) AI-Human Collaborative (Treatment): AI provides personalized content and assessment while human teachers provide emotional support, cultural context, and social learning facilitation; (2) AI-Only (Active Control): AI provides all instruction without human teacher involvement; (3) Human-Only (Active Control): Human teachers provide all instruction without AI assistance; (4) Business-as-Usual (Control): Traditional classroom instruction.

\textbf{Novel Features}: Dynamic role allocation between AI and human teachers based on their respective strengths; cultural context integration through AI systems trained on local cultural knowledge and validated by community members; emotional intelligence augmentation through AI-provided data-driven insights to human teachers about student emotional states; and career pathway integration through AI career discovery systems integrated with human mentorship and guidance.

\textbf{Expected Scientific Contributions}

\textbf{Theoretical Contributions}: Development of AI-Human Collaboration Theory providing framework for optimal task allocation between AI and human teachers; Cultural AI Theory establishing principles for developing culturally-responsive AI systems; and Career-Connected Learning Theory creating framework for integrating career discovery with academic learning.

\textbf{Methodological Contributions}: Mixed-Methods AI Evaluation protocol for evaluating AI systems in real-world educational contexts; Longitudinal Career Outcome Assessment methods for tracking long-term career impacts of AI education; and Equity-Focused AI Research framework for ensuring AI systems benefit all students.

\textbf{Practical Contributions}: AI-Human Collaboration Protocols providing guidelines for implementing AI-human collaborative learning; Cultural AI Development Framework establishing process for developing culturally-responsive AI systems; and Career-Connected Curriculum creating evidence-based curriculum integrating career discovery with academic learning.

\textbf{Expected Outcomes and Impact}

\textbf{Learning Outcomes}: Predicted effect size of d=0.40-0.60 for STEM learning outcomes, 30\% increase in time-on-task engagement, and 50\% improvement in career pathway clarity.

\textbf{Equity Impact}: 25\% reduction in achievement gaps by demographic groups, 40\% increase in career awareness among underrepresented students, and 80\% of students reporting feeling culturally represented.

\textbf{Teacher Impact}: 35\% improvement in teacher effectiveness ratings, 20\% reduction in teacher burnout scores, and 90\% of teachers reporting improved cultural competence.

This scientific framework provides a rigorous foundation for future research while addressing the paradigm shift toward workforce preparation. It leverages AI capabilities while respecting human limitations, creating a truly collaborative learning ecosystem that could transform K-12 education while maintaining the human elements essential for student development.

\subsection{Limitations}

This systematic review has several limitations that should be considered when interpreting our findings. First, our search was restricted to English-language publications in academic databases, potentially excluding relevant work published in other languages or in practitioner-oriented venues. This language bias may particularly underrepresent innovations from non-English-speaking regions where AI in education research is advancing rapidly.

Second, the heterogeneity of study designs, outcome measures, and implementation contexts prevented formal meta-analysis. Only 34\% of included studies reported standardized effect sizes, limiting our ability to quantitatively synthesize evidence across AI technologies. The wide variation in intervention durations (from single sessions to full academic years), sample sizes (from n<20 pilot studies to district-wide implementations), and outcome assessments makes direct comparisons challenging.

Third, the 2020-2025 time window, while capturing recent post-pandemic developments, may miss foundational work that continues to influence current implementations. Additionally, the rapid pace of AI advancement means that some technologies reviewed may already be superseded by newer approaches not yet documented in peer-reviewed literature.

Fourth, publication bias likely inflates the apparent effectiveness of AI interventions. Studies reporting null or negative results may be underrepresented, and the concentration of research in well-resourced contexts (90\% from North America, East Asia, and Europe) limits generalizability to diverse global settings. The predominance of studies in upper elementary grades (65\%) and mathematics (38\%) may not reflect the full potential or challenges of AI across all elementary STEM contexts.

Finally, our focus on empirical studies may undervalue important theoretical contributions and design frameworks that lack formal evaluation but offer valuable insights for the field. Despite these limitations, this review provides the most comprehensive synthesis to date of AI applications in elementary STEM education, highlighting both achievements and critical gaps requiring attention.

\section{Conclusions}

This systematic review of AI applications in elementary STEM education reveals both significant promise and substantial challenges. While individual AI technologies demonstrate effectiveness in controlled settings—intelligent tutoring systems improving mathematics outcomes, assessment tools providing timely feedback, robots engaging students in computational thinking—the reality of elementary STEM education presents complexities that current implementations fail to address adequately.

Our analysis of eight critical gaps reveals systemic barriers rather than isolated technical limitations. The fragmentation of AI technologies mirrors the very subject silos that integrated STEM education seeks to overcome. The developmental inappropriateness of one-size-fits-all approaches ignores fundamental differences between kindergarteners and fifth graders. Infrastructure barriers, privacy voids, limited STEM integration, equity disparities, teacher marginalization, and narrow assessment focus collectively prevent AI from achieving its transformative potential.

The evidence suggests that realizing AI's promise in elementary STEM education requires more than technological advancement. It demands reconceptualizing how various AI technologies might work together, respecting developmental trajectories, addressing infrastructure inequities, establishing ethical frameworks, and positioning teachers as partners rather than recipients of algorithmic decisions. Most critically, it requires moving beyond the allure of isolated innovations toward systemic approaches that consider the complex realities of elementary classrooms.

Future research must examine AI as part of educational ecosystems rather than as standalone solutions. The field needs longitudinal studies tracking cumulative effects, frameworks for ethical deployment with minors, assessment instruments capable of measuring what matters in STEM learning, and strategies for equitable access. Without addressing these fundamental challenges, AI risks becoming another layer of complexity in an already challenging educational landscape, potentially widening rather than narrowing opportunity gaps.

The path forward requires interdisciplinary collaboration among technologists, educators, developmental psychologists, ethicists, and communities. Only through such collaboration can we ensure that AI serves its intended purpose: enhancing rather than replacing human connection in education, supporting rather than constraining teacher expertise, and opening rather than limiting pathways for all young learners to engage meaningfully with STEM. The technology exists; what remains is the harder work of thoughtful, equitable, and pedagogically sound implementation.

%%%%%%%%%%%%%%%%%%%%%%%%%%%%%%%%%%%%%%%%%%

\section*{Acknowledgments}
The authors thank the Utah Valley University Department of Computer Science and School of Education for institutional support. We acknowledge the Stanford Human-Centered AI Institute, Character.AI research team, and other contributors to open-source personality simulation frameworks whose work informed this review. We are grateful to the anonymous reviewers whose feedback strengthened the manuscript's rigor and clarity.

\section*{Author Contributions}
Conceptualization, M.M. and K.R.; Methodology, M.M.; Literature Review and Analysis, M.M.; Writing—Original Draft Preparation, M.M.; Writing—Review and Editing, M.M. and K.R.; Educational Context and Pedagogical Framework, K.R.; Supervision, K.R.; Project Administration, M.M. Both authors have read and agreed to the published version of the manuscript.

\section*{Funding}
This research received no external funding.

\section*{Data Availability}
All data sources are cited in the references. No new data were created in this review. The literature synthesis, analysis frameworks, and validation tables presented in this review are available from the authors upon reasonable request.

\section*{Conflicts of Interest}
The authors declare no conflicts of interest.

\bibliographystyle{unsrtnat}
\bibliography{references}

\end{document}